\documentclass[12pt]{article}

\usepackage{a4,amsmath,epsfig}
\usepackage{axodraw}

\newcommand{\nn}{\nonumber}
\newcommand{\eqn}[1]{Eq.(\ref{#1})}
\newcommand{\fig}[1]{Fig.\ref{#1}}
\newcommand{\tab}[1]{Table~\ref{#1}}

\newcommand{\sss}[1]{\mbox{\scriptsize #1}}

\newcommand{\LL}{{\cal L}}
\newcommand{\real}{{\cal\mbox{Re\,}}}
\newcommand{\imag}{{\cal\mbox{Im\,}}}

\newcommand{\M}{{\cal M}}
\newcommand{\SL}{{\cal S}_{\sss{L}}}
\newcommand{\SNL}{{\cal S}_{\sss{NL}}}

\newcommand{\A}{{\cal A}}
\newcommand{\V}{{\cal V}}

\newcommand{\T}{\mbox{T}}

\newcommand{\Tr}{\mbox{Tr}\,}
\newcommand{\TF}{\mbox{\boldmath $F$}}

\newcommand{\TW}{\mbox{\boldmath $W$}}

\def\sj{\sin\theta}
\def\cj{\cos\theta}

\newcommand{\GeV}{\unskip\,\mathrm{GeV}}

\newcommand{\MW}{M_{_W}}
\newcommand{\MZ}{M_{_Z}}
\newcommand{\muW}{\mu_{_W}}
\newcommand{\muZ}{\mu_{_Z}}
\newcommand{\sw}{s_{_W}}
\newcommand{\cw}{c_{_W}}
\newcommand{\sumf}{\sum\limits_f N_{_C}^f\,}
\newcommand{\sumperm}{\sum\limits_{\sss{perm}}}
\newcommand{\pifs}{\Pi_f^{\gamma}(s)}
\newcommand{\sigone}{\tilde{\Sigma}_1}
\newcommand{\sigtwo}{\tilde{\Sigma}_2}
\newcommand{\sigthree}{\tilde{\Sigma}_3}
\newcommand{\sigfour}{\tilde{\Sigma}_4}
\newcommand{\sigfive}{\tilde{\Sigma}_5}
\newcommand{\sigsix}{\tilde{\Sigma}_6}
\newcommand{\Sone}{\tilde{S}_1}
\newcommand{\Stwo}{\tilde{S}_2}
\newcommand{\Sthree}{\tilde{S}_3}
\newcommand{\Sfour}{\tilde{S}_4}
\newcommand{\Sfive}{\tilde{S}_5}
\newcommand{\Ssix}{\tilde{S}_6}
\newcommand{\fun}{G}
\newcommand{\gbbc}{g_{_{\sss{\BBC}}}}

\def\gmk{g^{\mu\kappa}}
\def\gml{g^{\mu\lambda}}
\def\gkl{g^{\kappa\lambda}}
\def\ppm{p_+^\mu}
\def\pmm{p_-^\mu}
\def\ppk{p_+^\kappa}
\def\pmk{p_-^\kappa}
\def\ppl{p_+^\lambda}
\def\pml{p_-^\lambda}

\def\qpp{(qp_+)}
\def\qpm{(qp_-)}
\def\pps{p_+^2 }
\def\pms{p_-^2 }
\def\pppm{(p_+p_-)}
\def\form{Form }
\def\BBC{BBC }
\def\FL{FL }
\def\FW{FW }
\def\BBCN{BBCN }
\def\feyncalc{FeynCalc }
\def\WI{Ward Identities }

\begin{document}

\title{
{ \normalsize \begin{flushright}
                CERN-TH/2003-052 \\
                hep-ph/0303105
              \end{flushright} \vspace{1.5cm} 
}
Towards an effective-action approach to fermion-loop corrections
\footnote{Work supported by the European Union 
under contract HPRN-CT-2000-00149}}
\author{W.~Beenakker$^1$, A.P.~Chapovsky$^2$, A.~Kanaki$^3$,\\ 
C.G.~Papadopoulos$^{3,4}$\footnote{Financially supported
by EU contract HPMF-CT-2002-01622}, R.~Pittau$^5$\footnote{Financially supported
by MIUR under contract 2001023713-006}}
\date{}
\maketitle
\vspace{0.5cm}
\begin{center}
{\small
$^1$ Theoretical Physics, Univ. of Nijmegen,
NL-6500 GL Nijmegen, The Netherlands
\\
$^2$ Institut f\"ur Theoretische Physik E, RWTH Aachen, D-52056 Aachen, Germany
\\
$^3$ Institute of Nuclear Physics, NCSR ``Demokritos'', 15310, Athens, Greece
\\
$^4$ Theory Division, CERN, CH-1211, Geneva, Switzerland
\\ 
$^5$ Dipartimento di Fisica Teorica, Universit\`a di Torino, Italy \\ 
and INFN sezione di Torino, Italy}
\end{center}
\vspace{0.5cm}
\begin{abstract}
We present a study of the effective action approach to incorporate 
higher-order effects in $e^+e^-\to n\mbox{ fermions}$. In its minimal version,
the effective action approach is found to exhibit problems with 
unitarity and high-energy behaviour. We identify the origin of these problems
by investigating the 
zero-mode solutions of the Ward Identities. A numerical analysis of the 
importance of the zero-mode solutions is presented for four-fermion production 
processes.
\end{abstract}

\newpage

\section{Introduction}

Multi-fermion production processes constitute one of the most important classes
of reactions at electron--positron colliders \cite{epem}. 
Through high-precision studies of these 
reactions valuable information is gained on the electroweak parameters, on the
interactions between the electroweak gauge 
bosons and on the mechanism of electroweak symmetry breaking.   
High-precision studies of this kind demand a 
precise description of the physics of the unstable gauge bosons that occur 
during the intermediate stages of the reactions. 
One problematic, though crucial ingredient for achieving such a  
description is the incorporation of the associated finite-width effects.
To this end one has to resum the relevant gauge-boson
self-energies, which results in a mixing of different orders of perturbation 
theory and thereby jeopardizes gauge invariance. Since the high precision of 
the experiments has to be matched by the precision of the theoretical 
predictions, both an adequate treatment of the finite-width effects and a 
sufficiently accurate perturbative expansion are required. The clash between
resummation and perturbative expansion can therefore not be ignored. 

A procedure to overcome this dilemma has been proposed several 
years ago and is known under the name of Fermion Loop (\FL) scheme 
\cite{Argyres:1995ym,Beenakker:1996kn,Passarino:1999zh,Accomando:1999zq}. In this 
scheme a resummation of all one-loop fermionic corrections to gauge-boson 
self-energies is performed. In order to account for a consistent and 
gauge-invariant treatment, the one-particle-irreducible (1PI) fermionic 
one-loop corrections to the other $n$-point gauge-boson functions 
(with $n\ge 3$) are included as well. The \FL scheme essentially involves the
closed subset of all ${\cal O}([N_{_C}^f\alpha/\pi]^n)$ contributions to a 
given physical process, with $N_{_C}^f$ denoting the colour degeneracy of 
fermion $f$, 
and as such it is manifestly consistent. The reason for singling out the 
fermionic one-loop corrections lies in the fact that the unstable gauge bosons 
decay exclusively into fermions at lowest order. The \FL scheme has proven 
particularly successful in dealing with four-fermion production processes. 
Although in the beginning it merely served the purpose of a consistent scheme 
for including the width of the $W$ boson \cite{Argyres:1995ym}, which is 
closely related to the imaginary part of the $W$-boson self-energy, very soon 
people realized that it can also accommodate the resummation of the real parts 
of the gauge-boson self-energies \cite{Beenakker:1996kn,Passarino:1999zh}, 
which are responsible for the running of the couplings with energy.

Unfortunately there are several limitations related to the \FL scheme. First of
all, it is clearly a partial answer to the problem of resumming higher-order
corrections. It is restricted to closed fermion loops, which means that 
bosonic contributions are ignored. Several methods have been proposed to 
overcome this limitation. The most efficient one is the so-called 
pole-scheme \cite{Stuart:1991xk}, which amounts to a systematic expansion of 
the matrix elements 
around the complex poles in the unstable-particle propagators. In leading order
of this expansion the radiative corrections involve the {\it full} set of 
one-loop corrections to on-shell gauge-boson production and decay (factorizable
corrections) \cite{Beenakker:1998gr,Denner:2000bj}, as well as soft-photonic 
corrections that take into account the fact that the production and decay 
stages of the reaction do not proceed independently (non-factorizable 
corrections) \cite{nonfact}. 
However, in reactions with several intermediate unstable gauge 
bosons, like e.g.~six-fermion production, it becomes rather awkward to perform 
the complete pole-scheme expansion \cite{Beenakker:1998gr}. 
Secondly, even though the \FL scheme is conceptually straightforward, it 
becomes more and more involved computationally once one goes beyond the 
four-fermion production processes. For instance, for general multi-fermion 
production processes one has to consider the complete set of fermionic one-loop
corrections to the 1PI four-point gauge-boson functions, five-point gauge-boson
functions, and so on.  

In the meantime a novel proposal has emerged, as described in the paper by 
Beenakker, Berends and Chapovsky \cite{Beenakker:1999hi}, abbreviated as \BBC 
from now on. Their proposal consists essentially in a re-arrangement of the 
expansion of the effective action of the theory, which is usually performed in 
terms of the 1PI Feynman amplitudes, in such a way that the new expansion
is manifestly gauge invariant. Restricting ourselves, for simplicity, to a pure
$SU(N)$ gauge theory, the expansion looks like 
\begin{align}
\label{SNL}
{\cal S}_{NL} & = \int d^4x\, d^4y\, \fun_2(x,y)\, 
                  \Tr[ U(y,x) \TF_{\mu\nu}(x) U(x,y) \TF^{\mu\nu}(y) ]
\nn \\
              & + \int d^4x\, d^4y\, d^4z\, \fun_3(x,y,z)\,
                  \Tr[ U(z,x) \TF_{\mu\nu}(x) U(x,y) \TF^{\mu\rho}(y)
                       U(y,z) \TF^\nu_{\ \rho}(z) ]
\nn \\ 
              & + \ldots
\end{align}
Here the trace has to be taken in group space and 
$\,\TF_{\mu\nu}\equiv \frac{\displaystyle{i}}{\displaystyle{g}}\,
[D_{\mu},D_{\nu}]\,$ is the $SU(N)$ field-strength tensor, 
expressed in terms of the covariant derivative $D_{\mu}$ and the gauge 
coupling $g$. The operator $\,U(x,y)\,$ is 
a path-ordered exponential, which carries the gauge transformation from one 
space--time point to the other (see Section~2 for a more detailed 
definition). In \eqn{SNL} each gauge-invariant non-local operator is multiplied
by an appropriate space--time function $\fun_i\,$ that can, in principle, be
computed within perturbation theory. In the context of fermionic loop effects,
the various terms in \eqn{SNL} can be viewed as the result of integrating out 
the fermions in the functional integral, resulting in a kind of non-local 
 lagrangian for gauge-boson interactions. The minimum number of gauge 
bosons that participate in the effective interaction is two for the first term 
of \eqn{SNL}, three for the second term, and so on. Note, however,
that each term will also generate all higher $n$-point interactions, 
through the expansion of the path-ordered exponentials (see Section~2).
These higher $n$-point interactions are essential for achieving gauge 
invariance for the individual terms of \eqn{SNL}. Since all ingredients for 
the resummation of the gauge-boson self-energies are contained in the first 
(self-energy-like) term of \eqn{SNL}, it was proposed in the \BBC approach to 
truncate the series at this first term. In this way an economic gauge-invariant
framework for resumming self-energies is obtained, leading to matrix elements 
that satisfy all relevant \WI\!\!. Two questions remain open at this point: 
``\,How should one match the space--time function $\fun_2$ with the actual 
fermion-loop corrections?\,'' and ``\,Is gauge invariance sufficient for 
obtaining well-behaved matrix elements?\,''.    

In this paper we undertake the effort to confront the \BBC idea with actual 
calculations, addressing in this way the two outstanding questions. 
In Section~2 we consider the matching aspect. We introduce the set of 
gauge-invariant operators that is relevant for an exact description of 
fermion-loop corrections in the two-point gauge-boson sector of the Standard 
Model (SM), involving  both electroweak gauge bosons and Higgs fields. 
In Section~3 we identify and analyse a problem with the high-energy behaviour 
of the matrix element for the reaction $e^+e^-\to W^+W^-$. This problem is 
related to the non-unitary character of the truncation in the \BBC approach.
We pin-point the source of the problem to be in the zero-mode solutions of 
the \WI\!\!, like the second term of \eqn{SNL}, which are absent in the 
\BBC approach.
In Section~4 the set-up of the calculations as well as the numerical results 
are presented and discussed. Particular emphasis is put on an investigation of
the numerical importance of the zero-mode solutions. 
Finally, the paper is concluded with a few appendices, where all relevant 
information pertaining to the non-local Feynman rules, renormalization schemes 
and the unitarity problem in the reaction $e^+e^-\to W^+W^-$ can be found.

\section{The effective-action approach}

\subsection{Notation and conventions}

Before turning our attention to the non-local lagrangian, we first
introduce the notation and conventions that will be used throughout the 
remainder of this paper. In the SM there are four gauge fields, 
the $SU(2)_L$ (isospin) gauge fields $W_{\mu}^a\,$ ($a=1,2,3$)\, and the 
$U(1)_Y$ (hypercharge) gauge field $B_{\mu}$. The corresponding field-strength 
tensors are given by
\begin{equation}
  \TF_{\mu\nu} = \partial_{\mu}\TW_{\nu} - \partial_{\nu}\TW_{\mu}   
                 -i\,g_2\,[\TW_{\mu},\TW_{\nu}]\ ,
  \qquad
  B_{\mu\nu} = \partial_{\mu}B_{\nu} - \partial_{\nu}B_{\mu}~,
\end{equation}
using the shorthand notations
\begin{equation}
  \TF_{\mu\nu} \equiv \T^a\, F_{\mu\nu}^a\ ,
  \qquad
  \TW_{\mu} \equiv \T^a\, W_{\mu}^a~.
\end{equation}
The $SU(2)_L$ and $U(1)_Y$ gauge couplings are indicated by $g_2$ and $g_1$,
respectively, and the $SU(2)_L$ generators $\T^a$ can be expressed in terms of 
the standard Pauli spin matrices $\sigma^a\,$ ($a=1,2,3$) according to 
$\T^a=\sigma^a/2$. These generators obey the commutation relation
$[\T^a,\T^b]=i\,\epsilon^{abc}\,\T^c$, with the $SU(2)$ structure constant 
$\,\epsilon^{abc}\,$ given by 
\begin{equation}
  \epsilon^{abc} = \left\{ 
                   \begin{array}{cl}
                      {}+1 & \mbox{if $(a,b,c)=$ even permutation of (1,2,3)}\\
                      {}-1 & \mbox{if $(a,b,c)=$ odd permutation of (1,2,3)} \\
                      0    & \mbox{else}
                   \end{array}~. \right.
\end{equation}

\noindent The physically observable gauge-boson states are given by
\begin{equation}
\label{WZAvsWB}
  W_{\mu}^{\pm} = \frac{1}{\sqrt{2}}\,( W_{\mu}^1 \mp i\, W_{\mu}^2 )\ ,
  \qquad
  Z_{\mu} = \cw W_{\mu}^3 + \sw B_{\mu}\ ,
   \qquad
  A_{\mu} = \cw B_{\mu} - \sw W_{\mu}^3~,
\end{equation}
for the $W^{\pm}$ bosons, $Z$ boson and photon, respectively. 
Here $\,\cw=g_2/\sqrt{g_1^2+g_2^2}\,$ and $\,\sw = \sqrt{1-\cw^2}\,$ are the 
cosine and sine of the weak mixing angle. The electromagnetic coupling constant
can be obtained from $g_1$ and $g_2$ according to $\,e = \sqrt{4\pi\alpha} = 
g_1g_2/\sqrt{g_1^2+g_2^2}\,$. 

Since we want to discuss the entire gauge-boson sector, we also need to 
introduce the would-be Goldstone bosons $\phi^{\pm}$ and $\chi$ that are
intimately linked to the longitudinal degrees of freedom of the massive
$W^{\pm}$ and $Z$ gauge bosons. To this end we introduce the ($Y=1$) 
Higgs doublet
\begin{equation}
  \Phi(x) = \left( \begin{array}{c}
                     \phi^+(x) \\{}
                     [v+H(x)+i\,\chi(x)]/\sqrt{2}
                   \end{array} \right)\ 
\end{equation}
and the corresponding covariant derivative
\begin{equation}
  D_{\mu} = \partial_{\mu} - i\,g_2 \TW_{\mu} + i\,\frac{g_1}{2}\,B_{\mu}~.
\end{equation}
Here $\,v/\sqrt{2}\,$ is the non-zero vacuum expectation value of the Higgs 
field, yielding $\,\MW=vg_2/2\,$ and $\,\MZ = \MW/\cw\,$ for the masses of the 
$W$ and $Z$ bosons in this convention. 

A few more definitions are needed for the description of the fermionic 
corrections to the various self-energies in the gauge-boson sector of the SM.
A generic SM fermion will be indicated by $f$ and its isospin partner by $f'$.
The $SU(3)_C$ colour factor, mass, electromagnetic charge and isospin of the
fermion $f$ are denoted by $N_{_C}^f$, $m_f$, $eQ_f$ and $I_f^3$, respectively.
Finally, the hypercharge of the left-handed and right-handed fermions is 
denoted by $Y_f^L$ and $Y_f^R$, respectively.

\subsection{The non-local  lagrangian}

Following Ref.~\cite{Beenakker:1999hi} we introduce an effective action
that includes all relevant two-point interactions in the gauge-boson sector, 
involving both gauge-boson and Higgs fields. This {\it non-local} lagrangian 
can be written as
\begin{align}
\label{nlaction}
\SNL =   &
-\frac{1}{4} \int d^4x\, d^4y\,
\Sigma_1(x-y)\,  B_{\mu\nu}(x) B^{\mu\nu}(y)\   
\nn\\
&  -\frac{1}{2} \int d^4x\, d^4y\, \Sigma_2(x-y)\,
\Tr[U_2(y,x)\TF_{\mu\nu}(x)U_2(x,y)\TF^{\mu\nu}(y)]\  
\nn\\
&  -\frac{2}{v^2}\,\frac{g_1}{g_2}\int d^4x\, d^4y\,
\Sigma_3(x-y)\,[\Phi^\dag(x) \TF_{\mu\nu}(x) \Phi(x)]\,B^{\mu\nu}(y) \   
\nn\\
& -\frac{4}{v^4}\int d^4x\, d^4y\,
\Sigma_4(x-y)\,[\Phi^\dag(x) \TF_{\mu\nu}(x) \Phi(x)]\,
               [\Phi^\dag(y) \TF^{\mu\nu}(y) \Phi(y)]\  
\nn\\
&  +\int d^4x\, d^4y\, \Sigma_5(x-y)\,
[ D_\mu \Phi (x) ]^\dag\, U_2(x,y)U_1(x,y)\,D^\mu \Phi(y)\  
\nn\\
&  +\frac{2}{v^2} \int d^4x\, d^4y\, \Sigma_6(x-y)\,
 [\Phi^\dag(x) D_{\mu} \Phi(x)]^\dag\,
 [\Phi^\dag(y) D^{\mu} \Phi(y)]~. 
\end{align} 
A few comments and definitions are in order here. First of all, the arguments
of the non-local coefficients $\,\Sigma_1(x-y),\ldots,\Sigma_6(x-y)\,$ follow 
directly from translational invariance. Furthermore, the trace appearing in the
2nd term has to be taken in $SU(2)_L$ group space. Finally, the path-ordered 
exponentials for the $SU(2)_L$ and $U(1)_Y$ gauge groups are defined according 
to 
\begin{align}
U_2(x,y) & = P exp \Bigl[ {}-ig_2\int_x^y \TW_{\mu}(\omega)\, 
                          d \omega^{\mu} \Bigr] 
\nn \\[1mm]
U_1(x,y) & = P exp \Bigl[ {}+ig_1\,\frac{Y}{2}\int_x^y B_{\mu}(\omega)\, 
                          d \omega^{\mu} \Bigr]~, 
\end{align}
where $Y=1$ for the Higgs doublet and $d \omega^{\mu}$ is the element of 
integration along some path
$\Omega(x,y)$ that connects the points $x$ and $y$. According to 
Ref.~\cite{Beenakker:1999hi} the path is defined in such a way that it does 
not involve closed loops, i.e.~the null path $\Omega(x,x)$ always has zero 
length. Moreover, the choice of path should be such that it gives rise to 
path-ordered exponentials with specific properties under differentiation. 

\noindent Let us repeat the main points of the \BBC approach:
\begin{itemize}
\item The \BBC effective action in \eqn{nlaction} is gauge invariant by 
      construction.
\item Through the expansion of the path-ordered exponentials, the effective 
      action incorporates a set of higher 3-,4-,$\ldots,n$-point functions 
      that automatically satisfy the Ward Identities of the theory. A complete 
      set of three-point Feynman rules based on \eqn{nlaction} is given in 
      Appendix A. 
\item A set of unknown coefficients $\Sigma_1,\ldots,\Sigma_6$ is introduced.
\end{itemize}

There are several ways to determine the unknown coefficients. Some simplified
expressions, corresponding to existing ad hoc approximations for incorporating 
finite-width effects, have already been presented in 
Ref.~\cite{Beenakker:1999hi}. 
These expressions involve only a partial resummation of the fermionic 
corrections, in contrast to the full 1PI resummation that is performed in the 
\FL scheme. In this paper we investigate how the unknown \BBC coefficients can
be matched with the well-established two-point fermion-loop contributions in 
the SM. By doing so, we obtain an exact correspondence between  
the SM and the effective \BBC action for all reactions that involve at most 
two-point interactions among the gauge bosons. For reactions that involve 
interactions among three gauge bosons or more, the effective \BBC approach 
provides us with a minimal set of contributions that is required for satisfying
all relevant \WI\!\!. Although this approximation cannot be identical to that 
of the \FL scheme, one might anticipate that it provides a much more economic 
approach to multi-fermion production processes. After all, in the \FL scheme 
one has to perform a complete calculation of the SM $n$-point functions with 
three or more external gauge bosons, which constitutes a rather intensive and 
costly procedure. On the other hand, by truncating the non-local action at
`two-point order' several parts of the higher-order corrections are neglected. 
It is therefore important to understand to what extent one can trust such an 
approximation. 

\subsection{The matching procedure} 

In order to set up the framework of our studies, we present in this subsection
the matching procedure, i.e.~the determination of the non-local coefficients
$\Sigma_1,\ldots,\Sigma_6$. Using the knowledge of all two-point 
functions in the \FL scheme (see Appendix B), we can perform the first level 
of matching: mapping the unrenormalized self-energies directly onto the 
non-local coefficients. The second level of matching, between the so-obtained 
non-local matrix elements/cross sections and the explicit experimental 
observables, should take care of any necessary redefinition (renormalization)
of couplings and masses. 

As can be seen from Appendix B, we indeed need all six non-local operators in 
\eqn{nlaction} in order to match the six independent gauge-boson self-energies 
(after tadpole renormalization). The 1st, 2nd and 5th operators in 
\eqn{nlaction} are non-local extensions of terms in the local 
SM lagrangian. They take care of all UV-divergent terms present in the 
fermionic one-loop self-energies. The remaining three operators are higher 
dimensional (dim$\,>4$). The corresponding coefficients are finite, as expected
for a renormalizable theory, and can be viewed as non-local versions of the 
oblique $S$-, $T$- and $U$-parameters of Peskin and Takeuchi 
\cite{Peskin:1991sw}. These operators are required for achieving an explicit 
breaking of the global isospin symmetry among the $SU(2)$ gauge bosons, 
usually referred to as custodial $SU(2)$ symmetry \cite{Sikivie:1980hm}. 
After all, also the loop effects in the SM explicitly break this global 
symmetry as a result of hypercharge interactions and specific fermion-mass 
effects.

Below we list the results for the first level of matching of the coefficients 
$\sigone,\ldots,\sigsix$, which represent the Fourier transforms of the 
non-local coefficients $\Sigma_1,\ldots,\Sigma_6$ appearing in \eqn{nlaction}.
These results will be expressed in terms of the transverse and longitudinal 
self-energy functions $\,\Sigma_T^{V_1V_2}(s)$ and $\,\Sigma_L^{V_1V_2}(s)$, 
where $\,V_{1,2}=\gamma,Z,W\,$ and $s$ represents the square of the momentum at
which the self-energies are evaluated. The explicit expressions for these 
functions can be found in Appendix B in Eqs.(\ref{pifgamma})--(\ref{SigmaL}). 

The transverse, pure hypercharge coefficient $\sigone$ can be obtained through 
the relation
\begin{align}
  \sigone(s) &= \frac{1}{s}\,\biggl\{ \cw^2\Sigma_T^{\gamma\gamma}(s)
      + 2\sw\cw\Sigma_T^{\gamma Z}(s) 
      + \sw^2\Bigl[ \Sigma_T^{ZZ}(s) - \Sigma_L^{ZZ}(s) \Bigr] \biggr\}
      \nonumber \\
             &= \frac{1}{2}\,\sumf \Biggl[ \biggl(\frac{Y_f^L}{2\cw}\biggr)^2
      + \biggl(\frac{Y_f^R}{2\cw}\biggr)^2 \Biggr] \pifs~,
\end{align}
where the vacuum-polarization function $\,\pifs\,$ is defined in 
\eqn{pifgamma}. Since $\pifs$ is UV-divergent, the same must be true for 
$\sigone(s)$. Note also that this first coefficient is proportional to $g_1^2$.

The mixed hypercharge--isospin coefficient $\sigthree$ reads
\begin{align}
  \sigthree(s) &= \frac{1}{s}\,\biggl\{ \cw^2\Sigma_T^{\gamma\gamma}(s)
      + \frac{\cw}{\sw}\,(\sw^2-\cw^2) \Sigma_T^{\gamma Z}(s) 
      - \cw^2\Bigl[ \Sigma_T^{ZZ}(s) - \Sigma_L^{ZZ}(s) \Bigr] \biggr\}
      \nonumber \\
               &= \frac{1}{2}\,\frac{\cw}{\sw}\,\sumf 
      \biggl(\frac{I_f^3}{\sw}\biggr)\biggl(\frac{Y_f^L}{2\cw}\biggr)\pifs~,
\end{align}
which is finite because of the quantum-number identity $\sumf I_f^3 Y_f^L = 0$.
With our definition, involving the extra factor $g_1/g_2$ in \eqn{nlaction},
this coefficient is proportional to $g_2^2$.

The remaining two transverse, pure isospin coefficients $\sigtwo$ and 
$\sigfour$ are given by
\begin{equation}
  \sigtwo(s) = \frac{1}{s}\,\Bigl[ \Sigma_T^{WW}(s) - \Sigma_L^{WW}(s) \Bigr]
      \ = \ \frac{1}{2}\,\sumf \biggl(\frac{I_f^3}{\sw}\biggr)^2 \pifs
      - \sigfour(s) 
\end{equation}
and
\begin{align}
  \sigfour(s) &= \frac{1}{s}\,\biggl\{ \sw^2\Sigma_T^{\gamma\gamma}(s)
      - 2\sw\cw\Sigma_T^{\gamma Z}(s)
      + \cw^2\Bigl[ \Sigma_T^{ZZ}(s) - \Sigma_L^{ZZ}(s) \Bigr] 
      \nonumber \\
      & \hphantom{= \frac{1}{s}aA}
      {}- \Bigl[ \Sigma_T^{WW}(s) - \Sigma_L^{WW}(s) \Bigr] \biggr\}
      \nonumber \\[1mm]
              &= {}- \frac{\alpha}{24\pi\sw^2}\,\sumf 
      \Biggl\{ \biggl(1+\frac{2m_f^2}{s}\biggr) \Bigl[ B_0(s,m_{f'},m_f) 
                                                     - B_0(s,m_f,m_f) \Bigr]
      \nonumber \\[1mm]
              &  \hphantom{- \frac{\alpha}{24\pi\sw^2}}
      {}- \frac{4m_f^2(m_f^2-m_{f'}^2)}{s^2} \Bigl[ B_0(s,m_{f'},m_f) 
                                                  - B_0(0,m_{f'},m_f) \Bigr] 
      \Biggr\}~,
\end{align}
where the scalar two-point functions $B_0$ are defined in the usual 
way~\cite{Denner:kt}. The first coefficient is clearly UV-divergent and the 
second one clearly finite. Both coefficients are again proportional to $g_2^2$.

The coefficients $\sigthree$ and $\sigfour$ vanish at high energies 
($s \gg m_f^2$) and in the absence of doublet splitting ($m_{f'} = m_f$).
The former reflects the fact that there are only two independent self-energies
in the unbroken SM, whereas the latter indicates that there is no 
fermion-mass-induced custodial $SU(2)$ breaking if the fermions within an 
$SU(2)$ doublet have the same mass. 
For $s=0$ we obtain $s\sigthree(s) = s\sigfour(s) = 0$, which 
implies that we can match the transverse gauge-boson sector without the
explicit need for finite shifts of the gauge-boson masses. Such finite shifts 
will occur only in the longitudinal/scalar sector, as they should.

In the longitudinal/scalar sector we have two coefficients to match:
\begin{equation}
  \sigfive(s) = \left. 
      {}- \frac{\Sigma_L^{WW}(s)}{\MW^2}\right|_{\sss{no tadpole}}
      \ = \ \frac{1}{8\pi^2 v^2}\,\sumf m_f^2\,B_0(s,m_f,m_f)
      - \sigsix(s)
\end{equation}
\begin{align} 
  \sigsix(s)  &= \frac{\Sigma_L^{WW}(s)}{\MW^2} 
      - \frac{\Sigma_L^{ZZ}(s)}{\MZ^2} 
      \ = \ {}- \frac{1}{8\pi^2 v^2}\,\sumf m_f^2\,
      \biggl\{ B_0(s,m_{f'},m_f) \nonumber \\[1mm]
              &  \hphantom{=\ } 
      {}- B_0(s,m_f,m_f) - \frac{m_f^2-m_{f'}^2}{s}\,\Bigl[ B_0(s,m_{f'},m_f) 
      - B_0(0,m_{f'},m_f) \Bigr] 
      \biggr\}~,
\end{align}
where again the first coefficient is clearly UV-divergent and the second one 
clearly finite. As was to be expected, both coefficients are proportional to
$1/v^2$. The finite shifts of the gauge-boson masses at $s=0$ have been
absorbed into the non-local $T$-parameter (or $\rho$-parameter)
\begin{align}
  \sigsix(0) = \frac{\Sigma_L^{WW}(0)}{\MW^2} - \frac{\Sigma_L^{ZZ}(0)}{\MZ^2}
             = {}- \frac{1}{16\pi^2 v^2}\,\sumf m_f^2\,
               \Biggl[ 1\!-\!\frac{m_{f'}^2}{m_f^2-m_{f'}^2}\,
                             \log\biggl(\frac{m_f^2}{m_{f'}^2}\biggr)
               \Biggr].
\end{align}
Like $\sigthree$ and $\sigfour$, also $\sigsix$ vanishes at high energies 
($s \gg m_f^2$) and in the absence of doublet splitting ($m_{f'} = m_f$).

The explicit expressions for the resummed gauge-boson propagators in the 
covariant $R_{\xi}$ gauge can be found in Appendix A for both the transverse 
and longitudinal/scalar sectors.

\subsection{Running couplings}

In the sequel of this section we show explicitly how the introduction of 
running couplings leads to an effective description, where in analogy to the 
\FL scheme one 
just has to replace bare with running couplings in tree-order matrix elements 
in order to properly take into account the resummed fermionic corrections.  
As we have seen from the explicit expressions for the various non-local
coefficients, all six non-local coefficients are proportional to just one type
of bare coupling. In order to make the discussion of the running couplings 
easier, we therefore extract these couplings from the coefficients:
\begin{eqnarray}
  \sigone(s)  &=& g_1^2\,\Sone(s) \nonumber\\[2mm]
  \sigtwo(s)  &=& g_2^2\,\Stwo(s) 
  \quad , \quad \sigthree(s)\ =\ g_2^2\,\Sthree(s)
  \quad , \quad \sigfour(s)\ =\ g_2^2\,\Sfour(s) \nonumber\\[1mm]
  \sigfive(s) &=& \frac{1}{v^2}\,\Sfive(s) 
  \quad , \quad \sigsix(s)\ =\ \frac{1}{v^2}\,\Ssix(s)~,
\end{eqnarray}
with similar relations in coordinate space between $\Sigma_i(x-y)$ and 
$S_i(x-y)\,$ for $\,i=1,\ldots,6$. 

Upon closer investigation of \eqn{nlaction}  we notice that $\SNL$ 
contains exclusively the combinations $\,\Phi/v$, $g_1\,B\,$ or $\,g_2\,W$.
This means that the couplings can be absorbed into the field-definition.
For the corresponding local SM action we have
\begin{eqnarray}
  \label{SM/action}
  \SL &=& -\,\frac{1}{4\,g_1^2} \int d^4x\,g_1\,B_{\mu\nu}(x)\,
          g_1\,B^{\mu\nu}(x) 
          - \frac{1}{2\,g_2^2} \int d^4x\,\Tr\Bigl[ g_2\,\TF_{\mu\nu}(x)\,
          g_2\,\TF^{\mu\nu}(x) \Bigr] \nonumber \\[1mm]
      & & +\,v^2\,\int d^4x\,
          \frac{1}{v}\,\Bigl[D_{\mu} \Phi(x)\Bigr]^{\dagger}\,
          \frac{1}{v}\,D^{\mu}\Phi(x)~.
\end{eqnarray} 
The UV-divergences contained in the non-local coefficients $\,S_i(x-y)\,$ 
have the simple form  $\,S_i^{\sss{UV}}\,\delta^{(4)}(x-y)\,$ 
for $i=1,2,5$. Combining this with the local SM action, we end up with the
minimal renormalization requirement that $\,1/g_1^2+S_1^{\sss{UV}}\,$, 
$\,1/g_2^2+S_2^{\sss{UV}}\,$ and $\,v^2+S_5^{\sss{UV}}\,$
should become finite.

Having this in mind, we perform the re-diagonalization procedure in the 
transverse sector by introducing the running couplings
\begin{eqnarray}
\label{running1}
  \frac{1}{g_1^2(s)} &\equiv& \frac{1}{s\,g_1^2}\,\biggl[ s + 
          \Sigma_T^{\gamma\gamma}(s) + \frac{\sw}{\cw}\,\Sigma_T^{\gamma Z}(s)
          \biggl]\ =\ \frac{1}{g_1^2} + \Sone(s) + \Sthree(s) \nonumber\\[1mm]
  \frac{1}{g_2^2(s)} &\equiv& \frac{1}{s\,g_2^2}\,\biggl[ s + 
          \Sigma_T^{\gamma\gamma}(s) - \frac{\cw}{\sw}\,\Sigma_T^{\gamma Z}(s)
          \biggl]\ =\ \frac{1}{g_2^2} + \Stwo(s) + \Sthree(s) + \Sfour(s)
          \nonumber\\[3mm] 
  v^2(s)             &\equiv& v^2 + \Sfive(s)~,
\end{eqnarray}
the finiteness of which is consistent with the minimal renormalization 
requirement given above. From this a few more (finite) running quantities can 
be derived:
\begin{eqnarray}
\label{running2}
  \frac{1}{e^2(s)} &\equiv& \frac{1}{4\pi\alpha(s)}
          \ \equiv\ \frac{1}{g_1^2(s)} + \frac{1}{g_2^2(s)}
          \ =\ \frac{1}{e^2} + \Sone(s) + \Stwo(s) + 2\Sthree(s) + \Sfour(s)
          \nonumber\\[1mm]
  \sw^2(s) &\equiv& \frac{e^2(s)}{g_2^2(s)} \nonumber\\[1mm]
  \cw^2(s) &\equiv& 1-\sw^2(s) \ = \ \frac{e^2(s)}{g_1^2(s)} \nonumber\\[1mm]
  \MW^2(s) &\equiv& \cw^2(s)\,\MZ^2(s)\ \equiv\ \frac{1}{4}\,v^2(s)\,g_2^2(s)~.
\end{eqnarray} 

At this point the correspondence with the low-energy $S$-, $T$- and 
$U$-parameters of Peskin and Takeuchi \cite{Peskin:1991sw} can be made more 
explicit (see e.g.~Ref.~\cite{Hewett:1997zp}): 
\begin{eqnarray}
  S &=& 16\,\pi\,\frac{\sw^2\cw^2}{e^2}\,\lim_{s\to 0}\,\frac{1}{s}\,\biggl\{
        \Sigma_T^{ZZ}(s) - \Sigma_L^{ZZ}(s)
        + \frac{\cw^2-\sw^2}{\sw\cw}\,\Sigma_T^{\gamma Z}(s)
        - \Sigma_T^{\gamma\gamma}(s) \biggr\} \nonumber\\[1mm]
    &=& {}-16\,\pi\,\Sthree(0) \nonumber\\[3mm]
  T &=& \frac{v^2}{\alpha(0)\,v^2(0)}\,\biggl\{ \frac{\Sigma_L^{WW}(0)}{\MW^2} 
        - \frac{\Sigma_L^{ZZ}(0)}{\MZ^2} \biggr\}
        \ =\ \frac{1}{\alpha(0)\,v^2(0)}\,\Ssix(0) \nonumber\\[1mm]
  U &=& 16\,\pi\,\frac{\sw^2}{e^2}\,\lim_{s\to 0}\,\frac{1}{s} \biggl\{
        \Sigma_T^{WW}(s) - \Sigma_L^{WW}(s) 
        - \cw^2\,\Bigl[ \Sigma_T^{ZZ}(s) - \Sigma_L^{ZZ}(s) \Bigr]
        \nonumber\\[1mm]
   & & {}+ 2\sw\cw\,\Sigma_T^{\gamma Z}(s)
       - \sw^2\,\Sigma_T^{\gamma\gamma}(s) \biggr\} 
       \ =\ {}-16\,\pi\,\Sfour(0)~.
\end{eqnarray}

Next we want to verify that indeed all couplings have become running ones and 
that the propagator matrix in the transverse neutral sector has become 
diagonal. The easiest way to do this is by realizing that the complete matrix
element for a given reaction can be written in terms of subsets of matrix 
elements, each with a particular configuration of intermediate gauge bosons
and associated would-be Goldstone bosons. For the discussion of a particular 
intermediate gauge/Goldstone boson that carries a particular momentum, the 
relevant set of matrix elements can be represented by a propagator function 
that multiplies two distinct gauge/Goldstone-boson currents. In analogy to 
what was done above, the trick is now to explicitly pull out the coupling 
strength in these currents. For the $B$- and $W^a$-currents this amounts to
$\,J_B=g_1\,j_B\,$ and $\,J_{W^a}=g_2\,j_{W^a}$, respectively. Similarly a 
factor $1/v$ is pulled out in the would-be Goldstone-boson currents:
$J_{\chi} = j_{\chi}/v\,$ and $\,J_{\phi} = j_{\phi}/v$. 
Finally, we have to proof that the combination of propagator functions and
pulled-out coupling factors gives rise to running couplings and diagonal 
propagators.

Let us start with the transverse neutral sector, where we have to switch to 
the physical mass eigenstates [see \eqn{WZAvsWB}]:
\begin{eqnarray}
\label{JgammaZ}
  J_{\gamma} &=& \cw J_B - \sw J_{W^3} \ =\ e\,(\,j_B-j_{W^3}) 
                 \nonumber \\[1mm]
  J_{Z}      &=& \sw J_B + \cw J_{W^3} \ =\ \sw g_1\,j_B + \cw g_2\,j_{W^3}~.
\end{eqnarray} 
The generic amplitude structure for intermediate transverse neutral gauge 
bosons then reads
\begin{eqnarray}
  &&\hspace*{-7ex}\Bigl( J_{\gamma}^{\mu}\ J_{Z}^{\mu} \Bigr)
    \left( \begin{array}{cc}
             P_{T,\,\mu\nu}^{\gamma\gamma}(q) & 
             \!\!P_{T,\,\mu\nu}^{\gamma Z}(q) \\[1mm]
             P_{T,\,\mu\nu}^{\gamma Z}(q)     & 
             \!\!P_{T,\,\mu\nu}^{ZZ}(q)
           \end{array}
    \right)
    \left( \begin{array}{c}
             J_{\gamma}^{\,'\,\nu} \\
             J_{Z}^{\,'\,\nu}
           \end{array}
    \right) \ = \nonumber\\[3mm]
  &&\hspace*{-7ex}\Bigl( j_B^{\mu}\ j_{W^3}^{\mu} \Bigr)\!
    \left( \begin{array}{cc}
             e  & \sw g_1 \\
             -e & \cw g_2
           \end{array}
    \right)\!
    \left( \begin{array}{cc}
             P_{T,\,\mu\nu}^{\gamma\gamma}(q) & 
             \!\!P_{T,\,\mu\nu}^{\gamma Z}(q) \\[1mm]
             P_{T,\,\mu\nu}^{\gamma Z}(q)     & 
             \!\!P_{T,\,\mu\nu}^{ZZ}(q)
           \end{array}
    \right)\!
    \left( \begin{array}{cc}
             e       & -e \\
             \sw g_1 & \cw g_2
           \end{array}
    \right)\!
    \left( \begin{array}{c}
             j_B^{'\,\nu} \\
             j_{W^3}^{'\,\nu}
           \end{array}
    \right).
\end{eqnarray}
Using the propagator functions listed in Appendix A and the definitions
of the running couplings in Eqs.(\ref{running1}) and (\ref{running2}),
one can rewrite this product of matrices according to
\begin{eqnarray}
  &&\hspace*{-5ex}
    \left( \begin{array}{cc}
             e  & \sw g_1 \\
             -e & \cw g_2
           \end{array}
    \right)\!
    \left( \begin{array}{cc}
             P_{T,\,\mu\nu}^{\gamma\gamma}(q) & 
             \!\!P_{T,\,\mu\nu}^{\gamma Z}(q) \\[1mm]
             P_{T,\,\mu\nu}^{\gamma Z}(q)     & 
             \!\!P_{T,\,\mu\nu}^{ZZ}(q)
           \end{array}
    \right)\!
    \left( \begin{array}{cc}
             e       & -e \\
             \sw g_1 & \cw g_2
           \end{array}
    \right) \nn\\[3mm]
  &&\hspace*{-5ex} =\ 
    \left( \begin{array}{cc}
             e(s)  & \sw(s)\,g_1(s) \\
             -e(s) & \cw(s)\,g_2(s)
           \end{array}
    \right)\!
    \left( \begin{array}{cc}
              \bar{P}_{T,\,\mu\nu}^{\gamma\gamma}(q) & \!\!\!\!0 \\
              0 & \!\!\!\!\bar{P}_{T,\,\mu\nu}^{ZZ}(q)
           \end{array}
    \right)\!
    \left( \begin{array}{cc}
             e(s)           & -e(s) \\
             \sw(s)\,g_1(s) & \cw(s)\,g_2(s)
           \end{array}
    \right). \nonumber\\
\end{eqnarray} 
The diagonal transverse propagators are given by
\begin{eqnarray}
  \bar{P}_{T,\,\mu\nu}^{\gamma\gamma}(q) &=& {}-\frac{i}{s}\,\biggl( g_{\mu\nu}
           - \frac{q_{\mu}q_{\nu}}{q^2} \biggr) \nonumber\\[1mm] 
  \bar{P}_{T,\,\mu\nu}^{ZZ}(q)           &=& 
           \frac{g_2^2\,\cw^2(s)}{g_2^2(s)\,\cw^2}\,P_{T,\,\mu\nu}^{ZZ}(q) 
           \nonumber\\[1mm] 
                                         &=& {}-\frac{i}{s}\,\biggl( g_{\mu\nu}
           - \frac{q_{\mu}q_{\nu}}{q^2} \biggr)
           \biggl\{1 - \frac{g_2^2(s)}{\cw^2(s)}\,\Sthree(s)
                   - \frac{\MZ^2(s)}{s\,\rho(s)}\biggr\}^{-1}~, 
\end{eqnarray}
using the non-local $\rho$-parameter
\begin{equation}
  \rho(s) = \frac{v^2+\Sfive(s)}{v^2+\Sfive(s)+\Ssix(s)}
          = \frac{v^2(s)}{v^2(s)+\Ssix(s)}~.
\end{equation}
For the $W$-boson the generic amplitude structure reads
\begin{equation}
  j_W^{\mu}\,g_2\,P_{T,\,\mu\nu}^{WW}(q)\,g_2\,j_W^{'\,\nu} 
  = 
  j_W^{\mu}\,g_2(s)\,\bar{P}_{T,\,\mu\nu}^{WW}(q)\,g_2(s)\,j_W^{'\,\nu}~,
\end{equation}
with
\begin{eqnarray}
  \bar{P}_{T,\,\mu\nu}^{WW}(q) &=& \frac{g_2^2}{g_2^2(s)}\,
                                   P_{T,\,\mu\nu}^{WW}(q) \nonumber\\[1mm] 
                               &=& {}-\frac{i}{s}\,\biggl( g_{\mu\nu} 
           - \frac{q_{\mu}q_{\nu}}{q^2} \biggr)
           \biggl\{1 - g_2^2(s)\,\Sthree(s) - g_2^2(s)\,\Sfour(s)
                   - \frac{\MW^2(s)}{s}\biggr\}^{-1}~. \nonumber\\ 
\end{eqnarray}
So, indeed all couplings have been transformed into (finite) running couplings
and the effective propagators are diagonal and finite.
The complex poles of the diagonalized transverse gauge-boson propagators can
be obtained by solving the equations
\begin{eqnarray}
  s &=& 0 \nonumber\\[4mm]
  s &=& \muZ \ =\ \frac{\MZ^2(\muZ)/\rho(\muZ)}
                       {1-\frac{g_2^2(\muZ)}{\cw^2(\muZ)}\,\Sthree(\muZ)}
        \nonumber\\[1mm]
  s &=& \muW \ =\ \frac{\MW^2(\muW)}{1-g_2^2(\muW)\,\Sthree(\muW)
                                     -g_2^2(\muW)\,\Sfour(\muW)}~,
\end{eqnarray}
for the photon, $Z$ boson and $W$ boson, respectively.\\
In the longitudinal/scalar sector we get 
\begin{eqnarray}
  &&\hspace*{-4ex}\Bigl( j_B^{\mu}\ j_{W^3}^{\mu} \Bigr)\!
    \left( \begin{array}{cc}
              e & \sw g_1 \\
             -e & \cw g_2
           \end{array}
    \right)\!
    \left( \begin{array}{cc}
             \!P_{L,\,\mu\nu}^{\gamma\gamma}(q,\xi_{\gamma})\! & \!0\! \\
             \!0\! & \!P_{L,\,\mu\nu}^{ZZ}(q,\xi_{_Z})\!
           \end{array}
    \right)\!
    \left( \begin{array}{cc}
             e       & -e \\
             \sw g_1 & \cw g_2
           \end{array}
    \right)\!
    \left( \begin{array}{c}
             j_B^{'\,\nu} \\
             j_{W^3}^{'\,\nu}
           \end{array}
    \right) \nonumber\\[3mm]  
  && =\ j_Z^{\mu}\,\frac{g_2}{2\cw}\,P_{L,\,\mu\nu}^{ZZ}(q,\xi_{_Z})\,
    \frac{g_2}{2\cw}\,j_Z^{'\,\nu} \ =\ j_Z^{\mu}\,\frac{g_2}{2\cw}\,
    \biggl( \frac{-\,i\,q_{\mu}q_{\nu}}{q^2} \biggr)\,D_L^{ZZ}(s,\xi_{_Z})\,
    \frac{g_2}{2\cw}\,j_Z^{'\,\nu}~.\nonumber\\ 
\end{eqnarray}
In order to achieve this simplification, the $Z$-boson interactions were split
into an electromagnetic and isospin piece according to
\begin{eqnarray}
    J_{Z} &=& \sw g_1\,j_B + \cw g_2\,j_{W^3}
         \ =\ \frac{\sw^2-\cw^2}{2\sw\cw}\,e\,(\,j_B-j_{W^3}) 
              + \frac{g_2}{2\cw}\,(\,j_B+j_{W^3}) \nn\\[1mm]
          &\equiv& \frac{\sw^2-\cw^2}{2\sw\cw}\,J_{\gamma} 
                   + \frac{g_2}{2\cw}\,j_Z~.
\end{eqnarray}
The electromagnetic Ward Identity $\,q\cdot J_{\gamma}=0\,$ then takes 
care of all scalar electromagnetic interactions, leaving behind a pure isospin 
piece that has to be combined with the would-be Goldstone boson $\chi$. 
In the next step we combine the left-over $Z$-boson amplitude with the 
corresponding $\chi$-amplitudes:
\begin{eqnarray}
  &&\hspace*{-5mm}\Bigl( j_Z^{\mu}\ j_{\chi} \Bigr)
    \left( \begin{array}{cc}
             \MZ/v & 0 \\
             0     & 1/v
           \end{array}
    \right)\!
    \left( \begin{array}{cc}
             \!P_{L,\,\mu\nu}^{ZZ}(q,\xi_{_Z}) & P^{Z\chi}_{\mu}(q,\xi_{_Z})\! 
                                               \\[1mm]
             \!P^{\chi Z}_{\nu}(q,\xi_{_Z}) & P^{\chi\chi}(s,\xi_{_Z})\!
           \end{array}
    \right)\!
    \left( \begin{array}{cc}
             \MZ/v & 0 \\
             0     & 1/v
           \end{array}
    \right)\!
    \left( \begin{array}{c}
             j_Z^{'\,\nu} \\
             j'_{\chi}
           \end{array}
    \right) \nonumber\\[3mm]  
  &&=\ j_{\chi}\,\frac{1}{v}
    \biggl[ {}-i\,\frac{\MZ^2}{s}\,D_L^{ZZ}(s,\xi_{_Z}) 
            - 2\,i\,\MZ\,P^{Z\chi}(s,\xi_{_Z}) + P^{\chi\chi}(s,\xi_{_Z})
    \biggr]\,\frac{1}{v}\,j'_{\chi} \nonumber\\[3mm]
  &&=\ j_{\chi}\,\frac{1}{v(s)}\,\frac{i}{s}\,\rho(s)\,
    \frac{1}{v(s)}\,j'_{\chi}
    \ =\ j_{\chi}\,\frac{1}{v(s)}\,\frac{i}{s}\,
         \biggl\{ 1 + \frac{\Ssix(s)}{v^2(s)} \biggr\}^{-1}
    \frac{1}{v(s)}\,j'_{\chi}~.  
\end{eqnarray}
Here we have used the propagator functions listed in Appendix A, the running  
couplings as defined in Eqs.(\ref{running1}) and (\ref{running2}), and the 
two neutral-current \WI $\,q\cdot J_{\gamma}=0\,$ and 
$\,q\cdot J_{Z}=i\MZ J_{\chi}\,$ for an incoming momentum $q$. 
So, again we obtain running 
couplings. In a similar way we can combine the $W$-boson amplitudes in the 
longitudinal/scalar sector with the corresponding $\phi$-amplitudes, yielding 
the generic amplitude structure $j_{\phi}\,i/[s\,v^2(s)]\,j'_{\phi}$. 

If we would now use the unitary gauge in the massive gauge-boson sector, 
$\xi_{_{W/Z}} \to \infty$, all propagators involving would-be Goldstone bosons 
would vanish and the dressed $W$-boson and $Z$-boson propagators would become
\begin{eqnarray}
\label{unitary_gauge}
  P^{WW}_{\mu\nu}(q,\xi_{_W}) 
          \!\!\!&\stackrel{\xi_{_W}\to\,\infty}{-\!\!\!\longrightarrow}&\!\!\! 
          {}-i\,\frac{g_2^2(s)}{g_2^2}\,
          \biggl\{1 - g_2^2(s)\,\Sthree(s) - g_2^2(s)\,\Sfour(s) \biggr\}^{-1}
          \frac{g_{\mu\nu}\!-\!q_{\mu}q_{\nu}/{\cal W}(s)}{s-{\cal W}(s)}
          \nonumber \\[1mm]
  P^{ZZ}_{\mu\nu}(q,\xi_{_Z}) 
          \!\!\!&\stackrel{\xi_{_Z}\to\,\infty}{-\!\!\!\longrightarrow}&\!\!\! 
          {}-i\,\frac{g_2^2(s)\,\cw^2}{g_2^2\,\cw^2(s)}\,
          \biggl\{ 1 - \frac{g_2^2(s)}{\cw^2(s)}\,\Sthree(s) \biggr\}^{-1}
          \frac{g_{\mu\nu}\!-\!q_{\mu}q_{\nu}/{\cal Z}(s)}{s-{\cal Z}(s)}~,
\end{eqnarray}
where 
\begin{eqnarray}
\label{calW/Z}
 {\cal W}(s) &=& \frac{\MW^2(s)}
                      {1 - g_2^2(s)\,\Sthree(s) - g_2^2(s)\,\Sfour(s)}
                 \nonumber \\[1mm] 
 {\cal Z}(s) &=& \frac{\MZ^2(s)/\rho(s)}
                      {1 - \frac{g_2^2(s)}{\cw^2(s)}\,\Sthree(s)}~.
\end{eqnarray}
This of course leads to a huge reduction of the effective number of Feynman 
rules.%
\footnote{If all fermions would be massless or if there would be no doublet 
          splitting ($m_f = m_{f'}$), then 
          $\,\Sthree(s)=\Sfour(s)=\Ssix(s)=0\,$ and 
          $\,{\cal W}(s) = \MW^2(s) = \cw^2(s){\cal Z}(s)$. In that case we 
          reproduce the propagators of the so-called massive fermion-loop (MFL)
          scheme for a ``massless internal world'' \cite{Passarino:1999zh}.}
These resummed expressions in the unitary gauge are a suitable starting point 
for the second level of matching, i.e.~the renormalization, which is performed
explicitly in Appendix C.

\section{High-energy behaviour \& the zero-mode solutions}

Although \eqn{nlaction} describes a gauge-invariant action, there are other 
properties of the local theory (SM) that are not shared by the truncated
effective action. The most pronounced one is certainly unitarity and the 
related high-energy behaviour of the matrix elements. To exemplify this point,
we have computed the matrix elements 
$\,{\cal M}[\,e^+(p_1)e^-(p_2) \to W^+(p_+) W^-(p_-)\,]\,$ 
analytically. In Appendix~D it is shown that the matrix element for 
transversely polarized $W$ bosons and left-handed electrons exhibits an 
incorrect high-energy behaviour as a result of the presence of the factor
\begin{align}
  (\,p_+p_-)\,\left(\frac{\sigtwo(\pps)-\sigtwo(\pms)}{\pps-\pms}\right)
  \nn
\end{align}
in the non-local triple gauge-boson interaction. This factor clearly diverges
for large energies, unless $\sigtwo$ is a constant. 
Furthermore, as we will see 
in the next section, there is a rather important numerical discrepancy in the 
calculation of the cross section 
$\,\sigma(e^+e^-\to e^-\bar{\nu}_e u\bar{d}\,)\,$ in the extreme forward 
region, which is dominated by the exchange of nearly on-shell space-like 
photons.

It is obvious that any difference between the \BBC approach and the 
calculations in the \FL scheme must originate from the different treatment of 
the three-point vertices, since the two-point functions are identical
in both schemes. In order to understand the discrepancies in a more explicit 
way, let us define by the generic symbol $\Delta\Gamma$ 
the difference between a three-point vertex as computed
in the \FL scheme ($\Gamma_{\sss{\FL}}$) and the one in the \BBC approach 
($\Gamma_{\sss{\BBC}}$). 
In the case of the photon, for instance, we obtain 
\begin{eqnarray*}
\Delta\Gamma^{\mu\kappa\lambda}_{\gamma W^+W^-}(q,p_+,p_-)&=&
 \Gamma^{\mu\kappa\lambda}_{\sss{BBC},\,\gamma W^+W^-}
-\Gamma^{\mu\kappa\lambda}_{\sss{FL},\,\gamma W^+W^-}
\\
\Delta\Gamma^{\mu\kappa}_{\gamma W^+\phi^-}(q,p_+,p_-)&=&
 \Gamma^{\mu\kappa}_{\sss{BBC},\,\gamma W^+\phi^-}
-\Gamma^{\mu\kappa}_{\sss{FL},\,\gamma W^+\phi^-} 
\\
\Delta\Gamma^{\mu\lambda}_{\gamma \phi^+ W-}(q,p_+,p_-)&=&
 \Gamma^{\mu\lambda}_{\sss{BBC},\,\gamma \phi^+ W-}
-\Gamma^{\mu\lambda}_{\sss{FL},\,\gamma \phi^+ W-}
\\
\Delta\Gamma^{\mu}_{\gamma \phi^+\phi^-}(q,p_+,p_-)&=&
 \Gamma^{\mu}_{\sss{BBC},\,\gamma \phi^+\phi^-}
-\Gamma^{\mu}_{\sss{FL},\,\gamma \phi^+\phi^-}~. 
\end{eqnarray*}
The momenta and Lorentz indices of the incoming gauge bosons are denoted by
$(q,\mu)$ for the photon, $(p_+,\kappa)$ for the $W^+$ boson and
$(p_-,\lambda)$ for the $W^-$ boson, respectively. Similarly, the momenta of 
the incoming would-be Goldstone bosons $\phi^{\pm}$ are
given by $p_{\pm}$. 

Since all two-point functions are identical in the \FL and \BBC schemes, 
the above vertex quantities should satisfy a number of equations, 
namely \WI with all two-point functions switched off. These so-called
zero-mode equations can be
written as 
\begin{align}
q_\mu\Delta\Gamma^{\mu\kappa\lambda}_{\gamma W^+W^-}(q,p_+,p_-)=0
\nn\\
p_{+\kappa}\Delta\Gamma^{\mu\kappa\lambda}_{\gamma W^+W^-}(q,p_+,p_-)-
M_W\Delta\Gamma^{\mu\lambda}_{\gamma \phi^+ W^-}(q,p_+,p_-)=0
\nn\\
p_{-\lambda}\Delta\Gamma^{\mu\kappa\lambda}_{\gamma W^+W^-}(q,p_+,p_-)+
M_W\Delta\Gamma^{\mu\kappa}_{\gamma W^+ \phi^-}(q,p_+,p_-)=0
\nn\\
q_\mu\Delta\Gamma^{\mu\kappa}_{\gamma W^+\phi^-}(q,p_+,p_-)=\,
q_\mu\Delta\Gamma^{\mu\lambda}_{\gamma \phi^+W^-}(q,p_+,p_-)=0
\nn\\
p_{+\kappa}\Delta\Gamma^{\mu\kappa}_{\gamma W^+\phi^-}(q,p_+,p_-)-
M_W\Delta\Gamma^{\mu}_{\gamma \phi^+ \phi^-}(q,p_+,p_-)=0
\nn\\
p_{-\lambda}\Delta\Gamma^{\mu\lambda}_{\gamma \phi^+W^-}(q,p_+,p_-)+
M_W\Delta\Gamma^{\mu}_{\gamma \phi^+ \phi^-}(q,p_+,p_-)=0
\nn\\
q_\mu\Delta\Gamma^{\mu}_{\gamma \phi^+\phi^-}(q,p_+,p_-)=0~.\hspace*{-1.4ex}
\label{zeromodeeqs}
\end{align}
For the $Z$ boson one obtains a similar set of zero-mode equations. In that 
case, however, also the would-be Goldstone boson $\chi$ will feature 
explicitly in the expressions. 

In order to study the zero modes in detail, we introduce the following 
general form of the triple gauge-boson vertex (excluding $\epsilon$-tensor 
contributions%
\footnote{It is well known that these terms satisfy the \WI 
          on their own, without involving the two-point functions.}):
\begin{align}
V^{\mu\kappa\lambda}_{\gamma W^+W^-}(q,p_+,p_-) 
    &= ig_{_{\gamma WW}}\,\Bigl\{ x_1\,\ppm\gkl + x_2\,\pmm\gkl + x_3\,\ppk\gml 
                          + x_4\,\pmk\gml \notag\\[1mm]
    &\hphantom{= ig_{_{VWW}}A}
                          + x_5\,\ppl\gmk + x_6\,\pml\gmk + x_7\,\ppm\pmk\pml 
                          + x_8\,\pmm\ppk\,\ppl \notag\\[2mm]
    &\hphantom{= ig_{_{VWW}}A} 
                          + x_9\,\ppm\ppk\pml + x_{10}\,\pmm\ppk\pml 
                          + x_{11}\,\ppm\pmk\ppl \notag\\[1mm]
    &\hphantom{= ig_{_{VWW}}A} 
                          + x_{12}\,\pmm\pmk\ppl+ x_{13}\,\ppm\ppk\ppl 
                          + x_{14}\,\pmm\pmk\pml \Bigr\}~,
\label{vertex}
\end{align}
where $\,g_{_{\gamma WW}}=e$.  
The coefficients $x_i$ are scalar functions that depend on the squared momenta 
and masses. As a result of CP-invariance, there is a general symmetry of this 
vertex under the simultaneous transformations
\begin{equation}
  p_+ \leftrightarrow -\,p_- 
  \qquad \mbox{and} \qquad
  \kappa \leftrightarrow \lambda~,
\label{symmetry}
\end{equation}
which turn incoming $W^{\pm}$ bosons into outgoing $W^{\pm}$ bosons with the 
same momenta and Lorentz indices. 
This results in the relation 
\begin{equation}
  x_i(q^2,p_+^2,p_-^2) \to -\,x_{s(i)}(q^2,p_-^2,p_+^2)~,
\label{symmetry_relations}
\end{equation}
where 
$\,s(i)=\{2,1,6,5,4,3,8,7,10,9,12,11,14,13\}\ $ for $\ i=\{1,\ldots,14\}$.

A similar Lorentz-covariant parametrization can be made for the other 
three-point vertices: 
\begin{align}
V^{\mu\kappa}_{\gamma W^+\phi^-}(q,p_+,p_-) 
    &= ig_{_{\gamma WW}}\,\Bigl\{ y_1\,\gmk + y_2\,\ppm\ppk+y_3\,\ppm\pmk 
                          + y_4\,\pmm\ppk + y_5\,\pmm\pmk \Bigr\}
\nn \\[1mm]
V^{\mu\lambda}_{\gamma \phi^+W^-}(q,p_+,p_-) 
    &= ig_{_{\gamma WW}}\,\Bigl\{ z_1\,\gml + z_2\,\ppm\ppl+z_3\,\ppm\pml 
                          + z_4\,\pmm\ppl + z_5\,\pmm\pml \Bigr\}
\label{gammaWphi}
\end{align}
and
\begin{align}
V^\mu_{\gamma \phi^+\phi^-}(q,p_+,p_-)=
ig_{_{\gamma WW}}\,w_1\Bigl[ (q\cdot p_-) p_+^\mu - (q\cdot p_+) p_-^\mu \Bigr]~,
\end{align}
where in the latter case the relevant Ward Identity has been taken 
into account. As a result of CP-invariance we may relate the coefficients 
$y_i$ to the coefficients $z_i$ in \eqn{gammaWphi}.

If  we now demand that all three-point 
vertices satisfy \eqn{zeromodeeqs}
we end up with 25 coefficients 
satisfying a system of 21 equations.
This can be solved algebraically in terms of 9 {\it coefficients}, the number 
of which can be reduced to 5 independent {\it functions} if the symmetry
relations are exploited. 
\clearpage

In order to keep the discussion of \eqn{zeromodeeqs} as simple as possible,
we choose to neglect all contributions from vertices involving would-be 
Goldstone bosons by considering exclusively massless fermions.
This can be done without loss of 
generality, since both $\sigtwo$ and the ensuing unitarity problem for 
transverse $W$ bosons are also present in the unbroken theory.  
In that case \eqn{vertex} is also valid for the $ZWW$ vertex, provided that 
$\,g_{_{\gamma WW}}$ is replaced by $\,g_{_{ZWW}}=-e\,\cw/\sw$.
It is not difficult to verify that the reduced system of zero-mode 
equations has  always a solution and that one can express all coefficients 
$x_i$ in terms of four independent ones. 
For instance, using $\,a=p_+^2$, $b=p_-^2$ and $c=(\,p_+p_-)$, 
a solution may be represented by  
\begin{align}
& x_1 = -\,\frac{b+c}{a-b}\left( a\,x_{11}+ b\,x_{12}-a\,x_{13} 
                                 -b\,x_{14} \right) 
\notag\\[1mm] 
& x_3 =    \frac{c\,(a+c)}{a-b}\left( x_{11}-x_{13} \right) 
         + \frac{c\,(b+c)}{a-b}\left( x_{12}-\frac{b}{a}\,x_{14} \right) 
\notag\\[1mm] 
& x_4 =    \frac{a\,(a+c)}{a-b}\left( -x_{11}+x_{13} \right)
         + \frac{b+c}{a-b}\,\left( -a\,x_{12}+b\,x_{14} \right) 
\notag\\[1mm] 
& x_7 =    \frac{a+c}{a-b}\,x_{11} 
         + \frac{b+c}{a-b}\,\left( x_{12}-\frac{a}{b}\,x_{13}-x_{14} \right) 
\notag\\[2mm] 
& x_9 = -\,\frac{c}{b}\,x_{13}~. 
\label{x1tox14}
\end{align}
The rest of the coefficients are determined by using the symmetry relations
(\ref{symmetry_relations}).
Notice that, although we have expressed the solution algebraically in terms of 
four {\it coefficients}, this number can be reduced to two independent 
{\it functions} by means of \eqn{symmetry_relations}.

The four algebraically independent Lorentz structures to be used in the 
zero-mode solution $\Delta\Gamma^{\mu\kappa\lambda}_{V W^+W^-}$ 
(for $V=\gamma,\,Z$) may be 
represented as follows in momentum space.
The simplest structure corresponds to 
$(x_{11},x_{12},x_{13},x_{14})=(b+c,-a-c,0,0)$ and reads
\begin{align}
 V_1^{\mu\kappa\lambda}
& =
\Bigl[ \qpm\ppm-\qpp\pmm \Bigr] \Bigl[ \pppm\gkl-\pmk\ppl \Bigr]~.
\label{V1}
\end{align}
The second one corresponds to the solution $(1,-1,0,0)$:
\begin{align}
 V_2^{\mu\kappa\lambda}
& = 
\Bigl[ \qpm\ppm-\qpp\pmm \Bigr]\gkl + \gml\Bigl[ \pppm\ppk-\pps\pmk \Bigr]
\nn\\
&-\gmk\Bigl[ \pppm\pml-\pms\ppl \Bigr]
-\ppm\pmk q^{\lambda}+\pmm q^{\kappa}\ppl~.
\label{V2}
\end{align}
Note that this vertex originates from the operator
\begin{align}
 {\cal O}_{FFF} = 
     \Tr[ U_2(z,x) \TF^{\mu}_{\ \nu}(x) U_2(x,y) \TF^{\nu}_{\ \sigma}(y)
          U_2(y,z) \TF^{\sigma}_{\ \mu}(z) ]~,\nn 
\end{align}
as was predicted in \eqn{SNL}. 
The third structure corresponds to $(0,0,b,-a)$:
\begin{align}
 V_3^{\mu\kappa\lambda}
& =
\Bigl[ \pms\gml-\pmm\pml \Bigr] \Bigl[ \qpp\ppk-\pps q^{\kappa} \Bigr]
-\Bigl[ \pps\gmk-\ppm\ppk \Bigr] \Bigl[ \qpm\pml-\pms q^{\lambda} \Bigr]~.
\label{V3}
\end{align}
Finally the fourth structure corresponds to 
$(0,0,b(b+c),-a(a+c))$:
\begin{align}
 V_4^{\mu\kappa\lambda}
& =
\Bigl[ \qpm\ppm-\qpp\pmm \Bigr]
\Bigl[ \pps\pms\gkl-\pps\pmk\pml-\pms\ppk\ppl+\pppm\ppk\pml \Bigr]~.
\label{V4}
\end{align}

The triple gauge-boson vertex in the \FL scheme, as presented in 
Ref.~\cite{Beenakker:1996kn}, can now be expressed in terms of the vertex in 
the \BBC approach plus a linear combination of all four zero modes of 
Eqs.(\ref{V1})--(\ref{V4})~\cite{aggelikiphd}. It is exactly this difference 
between the \BBC approach and the \FL scheme, i.e.~the zero-mode solution 
$\Delta\Gamma^{\mu\kappa\lambda}_{V W^+W^-}$, that we are after. 
For our purposes, however, it would be enough
to just determine the zero-mode solutions that apply to either the 
$\,q^2\uparrow 0\,$ or $\,q^2\to\infty\,$ limits, since in those limits the 
\BBC approach starts to deviate. 

There are several ways to attack the problem,
but we think that the most economical one would be to reduce as much as 
possible the information on the exact three-point vertex 
$\,\Gamma^{\mu\kappa\lambda}_{\sss{FL},\,V W^+W^-}$. 
This is motivated by the fact that we have future applications in mind where
vertices with more than three gauge bosons are needed, such as six-fermion 
processes or four-fermion processes with an additional photon. In those cases
one would like to avoid a complete fermion-loop computation as much as 
possible. In fact, we may further reduce the problem by taking into account the
fact that, at least for four-fermion processes, we are dealing with conserved
external currents. These conserved external currents are the result of either
having massless fermions in the final state or having massive fermions that 
couple to photons. This means that terms proportional to $q^{\mu}$, 
$p_+^{\kappa}$ and $p_-^{\lambda}$ can be neglected, leading to the following
simpler form for \eqn{vertex}:
\begin{eqnarray}
V^{\mu\kappa\lambda}_{V W^+W^-}(q,p_+,p_-)\!\!\!
      &=& \!\!\! ig_{_{VWW}}\Bigl\{ \frac{x_1\!-\!x_2}{2}\,(p_+-p_-)^\mu\,\gkl
          + \frac{x_{11}\!-\!x_{12}}{2}\,(p_+-p_-)^\mu\,\pmk\ppl \nn \\[1mm]
      & & \!\!\! \hphantom{ig_{_{VWW}}A}
          +\,x_4\,\pmk\gml + x_5\,\ppl\gmk \Bigr\}~. 
\label{rvertex} 
\end{eqnarray}
The idea is now to use the information from the triple gauge-boson vertex in 
the \FL scheme and keep only those terms that are proportional to the four  
tensor structures appearing in \eqn{rvertex}. The algebra of the vertex 
corrections has been performed with the help of \form\!\!~\cite{form}, 
resulting in an 
expression in terms of tensor coefficients~\cite{Denner:kt}. Subsequently, 
\feyncalc~\cite{feyncalc} 
has been used to reduce these tensor coefficients to scalar one-loop 
integrals according to the Passarino--Veltman decomposition. The results 
obtained in this way fully agree with the ones published in 
Ref.~\cite{Beenakker:1996kn}. In the next step, all terms proportional to the 
scalar three-point functions are discarded, since in the non-local approach we
consider only corrections based on two-point functions. A complete set of such 
three-point terms obviously satisfies the zero-mode equations, but it cannot 
compensate any incorrect high-energy behaviour originating from the two-point 
sector. The remaining expressions consist of terms proportional to the scalar
two-point functions $\,B_0(q^2,0,0)\,$ and $\,B_0(p_{\pm}^2,0,0)\,$ as well as 
rational terms that come from the tensor reduction and the four-dimensional
limit. Since our final goal is to provide a correction term to the \BBC 
description, it is more convenient to re-express these two-point functions in 
terms of the non-local coefficients $\,\sigtwo(q^2)\,$ and 
$\,\sigtwo(p_{\pm}^2)\,$ using the results of the previous section.
Subsequently, the fermion-mass dependence is restored in $\,\sigtwo$, which
will allow us to take the zero-virtuality limit.

Let us first discuss the final results for the zero-mode solution in the limit
$\,q^2\uparrow 0\,$. These results can be represented in the following way:
\begin{align}
\label{deltas}
  \delta_1 &\equiv   \bigl(\frac{x_1-x_2}{2}\bigr)_{\sss{\BBC}}\!\!
                   - \bigl(\frac{x_1-x_2}{2}\bigr)_{\sss{\FL}}\! =\ 0 
  \nn\\[2mm]
  \delta_2 &\equiv   \bigl(\frac{x_{11}-x_{12}}{2}\bigr)_{\sss{\BBC}}\!\!
                   - \bigl(\frac{x_{11}-x_{12}}{2}\bigr)_{\sss{\FL}}\!
            = {}- 16\,\gbbc\,\frac{s_2+s_3}{\left(s_2-s_3\right)^2} 
                - \frac{s_2+s_3}{s_2\,s_3}\;\sigtwo(s_1) 
  \nn\\[1mm]
           &\ \ \ {}+ \frac{2s_2^3-7s_2^2\,s_3+4s_2\,s_3^2-s_3^3}
                              {s_2\,\left(s_2-s_3\right)^3}\;\sigtwo(s_2)
                    - \frac{2s_3^3-7s_2\,s_3^2+4s_2^2\,s_3-s_2^3}
                              {\left(s_2-s_3\right)^3\,s_3}\;\sigtwo(s_3) 
  \nn\\[1mm]
  \delta_3 &\equiv   \bigl(x_4\bigr)_{\sss{\BBC}}\!\!
                   - \bigl(x_4\bigr)_{\sss{\FL}}\! 
            = 16\,\gbbc\,\frac{s_2}{s_2-s_3}
              + \frac{s_2}{s_3}\;\sigtwo(s_1)
  \nn\\[1mm]
           &\ \ \ {}+ \frac{s_2\,\left(-2s_2+3s_3\right)}
                              {\left(s_2-s_3\right)^2}\;\sigtwo(s_2)    
                    - \frac{s_2\,\left(s_2-2s_3\right)^2}
                              {\left(s_2-s_3\right)^2\,s_3}\;\sigtwo(s_3)
  \nn\\[1mm]
  \delta_4 &\equiv   \bigl(x_5\bigr)_{\sss{\BBC}}\!\!
                   - \bigl(x_5\bigr)_{\sss{\FL}}\!
            = {}-\delta_3(s_2 \leftrightarrow s_3)~,
\end{align}
where we have introduced the shorthand notations $\,s_1=q^2\,,\,s_2=p_+^2\,,\,
s_3=p_-^2\,$ as well as $\,\gbbc\!=g_2^2/(64\pi^2)\,$. The next step is to 
translate these four quantities into the basic coefficients $\,x_{11}$, $x_{12}$, 
$x_{13}\,$ and $\,x_{14}\,$ with the help of \eqn{x1tox14}:
\begin{align}
x_{11} &= \frac{\delta_2(s_1-s_2+s_3) - (\delta_3+\delta_4)}{s_1}
\nn \\[1mm]
x_{13} &= \delta_1\,\frac{s_2-s_3-s_1}{s_1 s_2}
        + \frac{s_3^2-(s_2-s_1)^2}{2s_1 s_2}\,
          \biggl[ \delta_2 - \frac{\delta_3+\delta_4}{s_3-s_2-s_1} \biggr]
\nn\\[1mm]
       &  \hspace*{2ex} {}+ \frac{2}{s_3-s_2-s_1}\,
          \Biggl[ \delta_3\,\frac{s_2+s_3-s_1}{2s_2} + \delta_4 \biggr]~,
\label{zeromodesol}
\end{align}
with $x_{12}$ and $x_{14}$ determined by means of \eqn{symmetry_relations}.
These four coefficients can be inserted in \eqn{x1tox14} in order to 
reconstruct a complete zero-mode solution that can be subtracted safely from 
the \BBC vertex. Note that the translation between 
$\,\delta_1,\ldots,\delta_4\,$ and $\,x_{11},x_{13}$, given in 
\eqn{zeromodesol}, has by itself already an important part of the information 
encoded. For instance, a finite difference between the \BBC and \FL vertex 
corrections in the limit $\,s_1\uparrow 0$ is equivalent with the conditions
\[ 
\delta_1=0 \qquad\mbox{and}\qquad \delta_2=\frac{\delta_3+\delta_4}{s_3-s_2}~,
\]
which is in full agreement with the explicit expressions in \eqn{deltas}. 
These conditions guarantee that the coefficients 
$\,x_{11},x_{12},x_{13},x_{14}\,$
are finite, which in turn guarantees that all $\,x_1,\ldots,x_{14}\,$ are 
finite, since no factors $\,1/s_1=1/(a+b+2c)\,$ are present in \eqn{x1tox14}.

The same exercise can be performed for the limit $\,s_1\to\infty$. In that case
we find 
\begin{align}
  \delta_1 &= {}- \frac{s_1}{2}\,\frac{\sigtwo(s_2)-\sigtwo(s_3)}{s_2-s_3}
                + 8\,\gbbc\! - \sigtwo(s_1)
                + \frac{s_2\,\sigtwo(s_2)-s_3\,\sigtwo(s_3)}{s_2-s_3}
  \nn \\[1mm]
  \delta_2 &= \frac{\sigtwo(s_2)-\sigtwo(s_3)}{s_2-s_3}
              + \frac{2\,\sigtwo(s_1)-\sigtwo(s_2)-\sigtwo(s_3)
                      -16\,\gbbc}{s_1}
  \nn \\[3mm]
  \delta_3 &= \delta_4 = 0~, 
\label{delta_he}
\end{align}
where we have kept all terms that can give rise to contributions to the
amplitude that are not suppressed by inverse powers of $s_1$. It is not 
difficult to see that the leading terms in \eqn{delta_he} can in fact be 
absorbed completely into the simplest zero-mode structure $V_1$ of \eqn{V1}, 
if multiplied by 
\[
\frac{ig_{_{VWW}}}{(\,p_+p_-)}\,
\biggl( \frac{\sigtwo(s_2)-\sigtwo(s_3)}{s_2-s_3}\biggr)~.
\]

This completes the explicit construction of the zero-mode solutions that should
contain the bulk of the differences between the \BBC approach and the SM in 
the limits $\,s_1\uparrow 0\,$ and $\,s_1\to\infty$. 
It is worthwhile to underline that the investigation performed in this section
does not, by any means, address the problem of unitarization 
of the effective \BBC action in general. The objective is to identify the 
differences, through the zero-mode solutions, between the \BBC and the \FL approach, 
which is manifestly unitary, in order to assess their physical significance. A 
numerical analysis of the importance of the zero-mode solutions is the subject of the 
next section.

\section{Results}

In this section we present, as an illustrative example of our approach, 
numerical results based on four-fermion production processes that involve 
interactions among three gauge bosons: the so-called CC20 and CC10 families. 
We focus our studies on three particular kinematical configurations:
\begin{enumerate}
\item the small-angle (or single-$W$) regime, using the process 
      $\,e^+e^-\to e^-\bar{\nu}_e u \bar{d}\,$ with a cut on the angle of the 
      outgoing electron;
\item the configuration without angular cuts, using the total cross section 
      for the process $\,e^+e^-\to e^-\bar{\nu}_e u \bar{d}\,$ 
      (which only involves technical cuts related to the use of massless 
      fermions);
\item the high-energy regime, using the process 
      $e^+e^-\to \mu^-\bar{\nu}_\mu u\bar{d}$.
\end{enumerate}

Our numerical analysis is based on {\tt NEXTCALIBUR}~\cite{nextcalibur}. 
The matrix-element computations are performed with the help of a new
version of {\tt HELAC}~\cite{helac} that includes all
relevant vertices coming from the 
non-local effective action of \eqn{nlaction}, as described in Appendix A. 
The gauge invariance of this implementation has been checked extensively 
by comparing the results for the 't\,Hooft--Feynman and unitary 
gauges. Particular attention has been paid to the numerical convergence
of the non-local coefficients $\tilde{\Sigma}_i$ in all possible ranges 
covered by both $q^2$ and the fermion masses. 
Finally, the computation of all necessary  one-loop
three-point tensor coefficient functions is based on the
numerical programme {\tt FF}~\cite{ff}.

The subtraction of the zero-mode solutions has been limited to the two
ranges $\,q^2\uparrow 0\,$ and $\,q^2 \to \infty$, where $q^2$ is the
virtuality of the relevant exchanged photon or $Z$ boson.
In the former limit $\,q^2 \equiv t=(p_e'-p_e)^2$, with
$p_e$ and $p_e'$ denoting the momenta of the incoming and 
the outgoing electrons. In the latter limit $\,q^2 \equiv s$, 
where $s$ represents the centre-of-mass energy squared of the process. 

In practice, one has to decide on the intervals of $q^2$ in which 
the zero-mode corrections are switched on. In the present calculation we 
have selected the range $\,-1\GeV^2\le q^2\le 0\,$ for the first kind of 
zero-mode correction%
\footnote{We have checked that our results remain the same when varying the 
          lower cut-off value between $\,-0.04\GeV^2\,$ and $\,-25\GeV^2$.}. 
For the high-energy regime we have applied the zero-mode corrections 
in the full $\,q^2\ge 0\,$ range, since our processes are anyway
dominated by double-resonant and single-resonant $W$-boson contributions.     

In \tab{t-ren} we summarize the input parameters of our renormalization 
scheme and give the resulting output values for the computed quantities. 
Typically, three bare quantities - the electromagnetic constant $e$, the 
weak coupling $g_2$ and the Higgs vacuum expectation value $v$ - 
have to be fixed by three experimental data points.
On the other hand, there are several well-measured experimental 
quantities. Therefore, in order to add part of the missing higher-order
contributions and improve the predictive power of our computation, 
we have decided to work with five experimental data points instead of three.
This means that, besides $e$, $g_2$ and $v$, two more parameters get fixed.
The first parameter is the top-quark mass $m_t$, 
which allows an effective description of the missing non-fermionic 
corrections at high mass scales~\cite{Argyres:1995ym}. 
The second parameter is the common light-quark mass, $m=m_u=m_d$, which allows 
us to take into account the electromagnetic constant at zero virtuality.
For the other fermionic masses we use their PDG values~\cite{pdg}.
The resulting running of the renormalized electromagnetic 
and weak couplings are presented in \fig{run_coup}.
More details on our renormalization procedure are given in Appendix C.
\begin{table}[ht]   
\begin{center} 
\begin{tabular}{|c||c|}  \hline 
       Input Parameters  &  Output values \\ 
\cline{1-2}
 & \\[-10pt]
   $m_{_W} = 80.35   \GeV$ & 
     $\sqrt{\real(\muW)} = 80.3235 \GeV $  \\
   $m_{_Z} = 91.1867 \GeV$ & 
     $-\,\frac{\imag({\displaystyle\muW})}{\sqrt{\real({\displaystyle\muW})}}
      = 2.0575 \GeV$  \\
   $\real[\alpha^{(5)}(m_{_Z}^2)^{-1}] = 128.89$ & 
     $m_u = m_d = 0.0475188 \GeV$ \\
   ${\alpha(0)}^{-1} = 137.03599976$ & 
     $m_t = 146.966 \GeV$  \\
   $G_F= 1.16639\times 10 ^{-5} \GeV^{-2}$ & 
   \\
\hline
\end{tabular}
\end{center}
\caption[.]{Input parameters versus computed quantities.}
\label{t-ren}
\end{table}
\begin{figure}[th]
\begin{center}
\mbox{
\epsfig{file=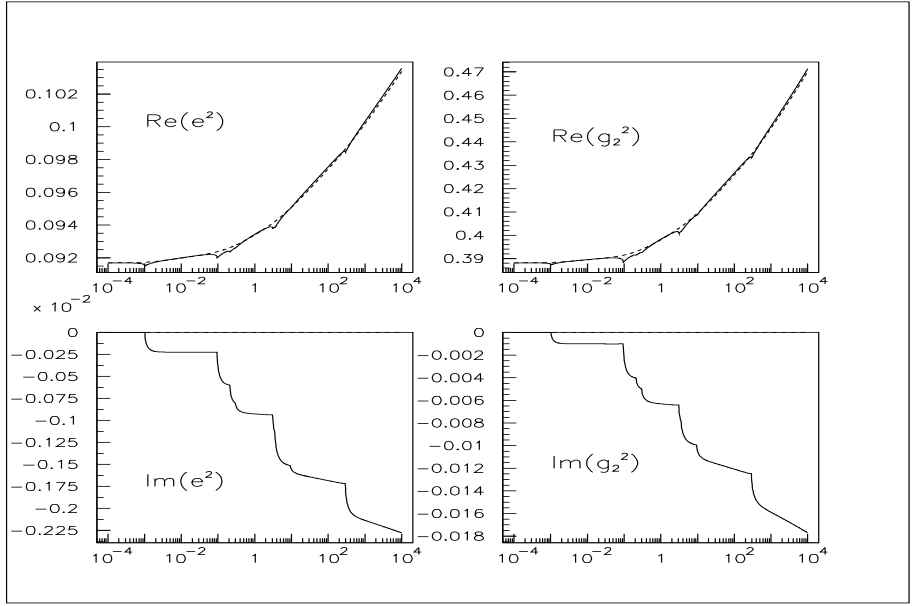,height=8cm,width=12cm}
}
\caption[.]{The evolution of the squared electromagnetic ($e^2$) and weak 
            ($g_2^2$) couplings as a function of the scale $|q|$ in 
            $\GeV$. The solid (dashed) line represents the evolution for 
            positive (negative) values of $q^2$. The values for $e^2$ and 
            $g_2^2$ predicted by the \FW scheme are given by 
            $0.09523$ and $0.4260$, respectively.}
\label{run_coup}
\end{center}
\end{figure}
 
Since it is our aim to compare different schemes, we now present a few 
different approaches. The first one is the widely used Fixed Width (\FW\!) 
scheme, where a fixed $W$-boson width is implemented in all
$W$-boson propagators and where the $G_F$-scheme is applied for evaluating 
the weak parameters. We recall that the latter is defined by using $m_{_W}$, 
$m_{_Z}$ and $G_F$ as input parameters, together with the two relations
\[
  \sw^2 = 1 - \frac{m_{_W}^2}{m_{_Z}^2} \qquad,\qquad
  \alpha = \frac{\sqrt{2}}{\pi}\,G_F\,m_{_W}^2 \sw^2~. 
\] 
In addition we introduce two hybrid schemes, where the
real (imaginary) part is fixed by the \FL (\BBC\!) scheme and vice versa. 
This we do in order to investigate possible differences between the real and 
imaginary parts of the corrections that are missing in the \BBC approach.
Finally, we denote by \BBCN the scheme that subtracts the relevant zero-mode 
solutions of the \WI\!\!.

In order to study the small-angle behaviour of the various approximations, 
we focus on the reaction $\,e^+ e^-\to e^-\bar{\nu}_e u\bar{d}\,$ with the 
following two cuts
\[ 
  |\cos(\theta_e)| > 0.997 \quad,\quad M(u\bar{d}\,) > 45 \GeV~. 
\]
The first cut, on the angle between the outgoing electron and the electron 
beam, ensures sensitivity to contributions that are mediated by $t$-channel 
graphs. The second cut, on the invariant mass of the $u\bar{d}$ system, is 
added mainly to comply with earlier calculations. The corresponding results 
are presented in \tab{t-satot}, from which we deduce that 
\begin{itemize}
\item the \FW scheme overestimates the cross sections by up to $6\%$. 
      This is mainly due to the use of non-running couplings, especially in
      the electromagnetic sector;
\item the \BBC approach underestimates the cross sections by up to $-6\%$.
      This, in contrast to the previous case, is due to differences 
      in the treatment of the triple gauge-boson vertex. This fact
      reflects the importance of subtracting zero-mode contributions;
\item the \BBCN approach reproduces the results obtained in the \FL scheme
      within MC accuracy.
\end{itemize}
\begin{table}
\begin{center}
\begin{scriptsize}
\begin{tabular}[t]{||c|c|c|c|c|c|c||}  \hline
$\!\!\sqrt{s}\ [\!\GeV]\!\!$ & \FW & \BBC 
& $\!\!\real(\mathrm{BBC})\!+\!\imag(\mathrm{FL})\!\!$
& $\!\!\real(\mathrm{FL})\!+\!\imag(\mathrm{BBC})\!\!$ & \BBCN  & \FL \\ 
\cline{1-7}  
  & & & & & & \\[-2mm]
  183 & 89.17(26) & 80.00(32) & 81.80(32) & 82.23(35) & 84.83(32)
  & 84.38(33) \\
  & & & & & & 83.28(6)\hphantom{1} \\
  189 & 99.80(24) & 89.38(34) & 92.19(35) & 92.02(35) & 95.13(36)
  & 94.60(36) \\
  & & & & & & 93.79(7)\hphantom{1} \\
  200 & 120.98(31) & 108.41(42) & 111.50(43) & 111.52(43) & 114.69(44)
  & 114.61(44)  \\
  & & & & & & 113.67(8)\hphantom{1} \\
  500 & 897.1(3.2) & 814.8(4.6) & 837.2(4.7) & 833.6(5.6) & 856.3(4.8)
  & 856.3(4.8) \\
  & & & & & & \\
  1000 & 2064(12) & 1931(16) & 1968(29) & 2042(55) & 1937(16) & 1964(16) \\
  & & & & & & \\
 \hline
\end{tabular}
\end{scriptsize}
\end{center}
\caption[.]{Cross sections (in $fb$) for the process 
            $\,e^+e^-\rightarrow e^-\bar{\nu}_e u\bar{d}$, using the cuts
            $\,|\cos(\theta_e)| > 0.997\,$ and $\,M(u\bar{d}\,) > 45\GeV$. 
            In each second row of \FL\!-scheme entries we give the results
            taken from Ref.~\cite{Passarino:2000mt}, which differ slightly
            from our results owing to a different treatment of the hadronic 
            part of the photonic vacuum polarization.}
\label{t-satot}
\end{table}
In \tab{t-sadif} the predicted cross section is shown for different angular 
regions of the outgoing electron. The four rows correspond to the \FW\!\!, 
\BBC\!\!, \BBCN and \FL schemes, respectively.
The previous observations, which were deduced for the total cross section 
with angular cut $\,|\cos(\theta_e)| > 0.997$, \,i.e.~$\theta_e < 4.44^o$, 
are more or less reproduced uniformly in the extreme-forward angular 
distribution between $0.0^o$ and $0.4^o$. 
\begin{table}
\begin{center}
\begin{tabular}[t]{||c|c|c|c||}  \hline
  $\theta_e$        & $183\GeV$ & $189\GeV$ & $200\GeV$ \\ \cline{1-4}
  $0.0^o$ - $0.1^o$ & 49.01(17) & 55.25(19) & 67.81(24) \\
                    & 42.98(23) & 48.55(26) & 59.35(32) \\
                    & 46.17(24) & 52.47(28) & 63.73(34) \\
                    & 45.56(24) & 51.53(27) & 63.17(34) \\ \cline{1-4}
  $0.1^o$ - $0.2^o$ & 7.03(7)   & 7.87(7)\hphantom{1} & 9.36(9)\hphantom{1} \\
                    & 6.16(9)   & 6.94(10)  & 8.35(12) \\
                    & 6.77(9)   & 7.63(11)  & 8.93(13) \\
                    & 6.60(9)   & 7.46(10)  & 8.84(13) \\ \cline{1-4}
  $0.2^o$ - $0.3^o$ & 4.21(5)   & 4.55(6)   & 5.40(7)\hphantom{1} \\
                    & 3.59(7)   & 4.21(8)   & 4.84(9)\hphantom{1} \\
                    & 3.92(7)   & 4.34(8)   & 5.17(10) \\
                    & 3.87(7)   & 4.26(8)   & 5.13(10) \\ \cline{1-4}
  $0.3^o$ - $0.4^o$ & 2.80(4)   & 3.23(5)   & 3.87(6) \\
                    & 2.61(6)   & 2.81(6)   & 3.55(8) \\
                    & 2.85(6)   & 3.08(7)   & 3.80(8) \\
                    & 2.81(6)   & 3.17(7)   & 3.76(8) \\  \hline
\end{tabular}
\end{center}
\caption[.]{Cross sections (in $fb$) for the process 
            $\,e^+e^-\rightarrow e^-\bar{\nu}_e u\bar{d}$, using the cut
            $\,M(u\bar{d}\,) > 45\GeV$. The results are presented for different
            energies $\sqrt{s}$ and for different bins of the angle $\theta_e$ 
            between the outgoing electron and the electron beam. The four rows 
            correspond to the \FW\!\!, \BBC\!\!, \BBCN and \FL schemes, 
            respectively.}
\label{t-sadif}
\end{table}

In \tab{t-tote} we present results {\it without} angular cuts.
The discrepancy is now reduced substantially, reflecting the
fact that an important component of the total cross section, 
namely the contribution of double resonant graphs, is 
equally well described by the different schemes.
This fact ceases to be true at energies above $500$ GeV where 
single-resonant and multi-peripheral contributions take over again. 
Nevertheless the \BBCN scheme still 
follows the \FL results within MC accuracy.

\begin{table}
\begin{center}
\begin{scriptsize}
\begin{tabular}[t]{||c|c|c|c|c|c|c||}  
\hline
$\!\!\sqrt{s}\ [\!\GeV]\!\!$ & \FW & \BBC 
& $\!\!\real(\mathrm{BBC})\!+\!\imag(\mathrm{FL})\!\!$
& $\!\!\real(\mathrm{FL})\!+\!\imag(\mathrm{BBC})\!\!$ & \BBCN  & \FL \\
\cline{1-7}  
  & & & & & & \\[-2mm]
  183 & 766.6(1.0) & 770.3(2.7) & 775.3(2.7) & 780.1(3.1) & 773.9(2.7)  
  & 777.7(3.1) \\[2mm]
  189 & 808.7(1.1) & 807.4(2.7) & 813.3(3.0) & 810.9(2.9) & 815.1(2.7)  
  & 814.2(3.0) \\[2mm]
  200 & 851.4(1.2) & 846.9(2.9) & 857.1(3.0) & 854.5(2.9) & 859.8(3.3)  
  & 860.9(3.0) \\[2mm]
  500 & 1377(4) & 1299(6) & 1344(8) & 1347(8) & 1341(6) & 1345(8) 
  \\[2mm]
  1000 & 2555(17) & 2387(16) & 2463(28) & 2471(30) & 2414(18) & 2463(27) 
  \\[2mm]
\hline\hline
\end{tabular}
\end{scriptsize}
\end{center}
\caption[.]{Cross sections (in $fb$) for the process 
            $\,e^+e^-\rightarrow e^-\bar{\nu}_e u\bar{d}$, using only the cut 
            $M(u\bar{d}\,) > 45\GeV$, \,i.e.~no angular cuts are imposed.}
\label{t-tote}
\end{table}

Finally, as already stated, the \BBC approach violates unitarity and as such 
gives rise to a bad high-energy behaviour. In order to better reveal this 
property, one has to go to rather high energies. To this end, we consider the
process $\,e^+e^-\to\mu^-\bar{\nu}_\mu u\bar{d}$. The results for the 
corresponding total cross section are presented in \tab{t-muon}, which shows
rather clearly that the \BBC approach and its hybrids start to diverge above 
$1$ TeV. In contrast, the \BBCN scheme exhibits a good high-energy behaviour.
\begin{table}
\begin{center}
\begin{scriptsize}
\begin{tabular}[t]{||c|c|c|c|c|c|c||}  \hline
$\!\!\sqrt{s}\ [\!\GeV]\!\!$ & \FW & \BBC 
& $\!\!\real(\mathrm{BBC})\!+\!\imag(\mathrm{FL})\!\!$
& $\!\!\real(\mathrm{FL})\!+\!\imag(\mathrm{BBC})\!\!$ & \BBCN  & \FL \\
\cline{1-7}
  & & & & & & \\[-2mm]
  200 & 686.86(81) & 702.8(2.5) & 704.6(2.4) & 704.6(2.4) & 702.3(2.5)  
  & 704.9(2.4) \\[2mm]
  500 & 270.71(45) & 275.5(1.2) & 276.4(1.2) & 276.5(1.2) & 275.0(1.2) 
  & 276.3(1.2) \\[2mm]
  1000 & 103.67(19) & 107.38(46) & 106.91(47) & 106.87(47) & 105.84(46)   
  & 106.10(47) \\[2mm]
  2000 & 36.107(75) & 43.36(18) & 40.05(18) & 40.10(19) & 36.45(17)   
  & 36.89(19) \\[2mm]
  5000 & 8.067(25) & 53.05(22) & 30.35(13) & 30.81(19) & 8.486(61)   
  & 8.225(45) \\[2mm]
  10000 & 2.445(11) & 187.53(62) & 94.66(39) & 95.31(44) & 4.227(26)   
  & 2.548(27) \\[2mm] \hline
\end{tabular}
\end{scriptsize}
\end{center}
\caption[.]{Cross sections (in $fb$) for the process 
            $\,e^+e^-\rightarrow \mu^-\bar{\nu}_\mu u\bar{d}$, using the cut  
            $M(u\bar{d}\,) > 45\GeV$.}
\label{t-muon}
\end{table}
However, in comparison with the \FW and \FL schemes, the \BBCN approach 
exhibits a substantial discrepancy above $\,\sim 7$ TeV. In trying to analyse 
this point, we found that the \BBCN and \FL schemes agree very well for 
massless fermions: e.g.~at $10$ TeV the \FL scheme gives $\sigma=1.209(77)$ fb
whereas the \BBCN approach yields $\sigma=1.207(77)$ fb. If we would use the 
nominal values for the masses of the light fermions, but reduce the top-quark 
mass to $\,m_t=10\GeV$, these numbers would change to 1.226(78) and 1.233(78), 
respectively. Recall that in the high-energy regime the \BBCN approach is 
defined by assuming massless fermions. The explicit fermion-mass dependence is 
re-introduced only through the non-local coefficients. The lesson to be 
learned here is that fermion masses, and more in particular the top-quark mass,
play a rather important role in the triple gauge-boson vertex and cannot be 
accommodated by the $\sigtwo$ non-local coefficient alone.

\section{Conclusions}

In this paper we have presented an analysis that represents a first step 
towards an effective-action description of fermion-loop corrections to 
multi-fermion reactions like $e^+e^-\to n\mbox{ fermions}$. 
The study is based on the proposal formulated in Ref.~\cite{Beenakker:1999hi}. 
It relies on a re-organization of the 
expansion of the effective action of the full theory. This re-organization is
performed in terms of gauge-invariant operators, involving an arbitrary number 
of gauge-boson and Higgs fields, multiplied by non-local coefficients. 
After a truncation of this expansion at the two-point-function level, 
a non-local effective theory is obtained that is consistent with all \WI for 
arbitrary $n$-point functions.

The next important step was the identification (matching) of the non-local 
coefficients with the fermionic one-loop self-energy contributions predicted 
within the Standard Model of electroweak interactions. Applied to physical 
processes that do not involve interactions among more than two gauge bosons,
the so-obtained effective theory and the Standard Model lead to identical
results. After renormalization, a complete description in terms of runnig 
couplings is established and the correct (fermion-loop) scale dependence is 
obtained, both in the high- and low-scale regimes.

Although the truncation of the expansion at the level of the two-point 
functions is gauge-invariant, it introduces nevertheless a bad high-energy 
behaviour by violating unitarity. This shows up, for instance, in the 
amplitudes for the production of {\it transversely polarized} massive gauge 
bosons, like $\,e^+e^-\to W^+_TW^-_T$. We have identified explicitly the zero 
modes of the \WI that are responsible for such behaviour. 
Based on the appropriate high- and low-scale limits, we have reconstructed 
the two simplest zero-mode solutions that, if subtracted from the non-local
triple gauge-boson vertex, would restore the agreement
between the effective theory and the Standard Model. 
More specifically, we have studied the numerical effect of the zero-mode 
solutions for several four-fermion production processes, both for CC20 and 
CC10 families. We have observed the following: 
\begin{itemize}
\item the effect of the zero modes is essential in restoring the
      good high-energy behaviour above $\sim 500\GeV$; 
\item the zero modes also account for the substantial discrepancy between the 
      effective theory and the Standard Model in the extreme forward region 
      if electrons/positrons are present in the final state; 
\item the contribution of the zero modes to the ``intermediate-energy''
      regime may be neglected safely.
\end{itemize}

It remains an open question how to construct in a more systematic way the 
relevant zero-mode solutions that will restore, to a given accuracy, 
the agreement between the effective theory and the Standard Model. 
Moreover, in order to extend the scheme to processes that involve interactions 
among four or more gauge bosons, like for instance six-fermion production in 
$e^+e^-$ collisions, special care should be devoted to a unitarity-preserving 
reformulation of the non-local effective action.
   
\vspace*{2cm}
{\bf\large Acknowledgments}

We would like to thank Frits Berends for his support during
the earlier stages of this work.

\newpage
\noindent {\large\bf Appendix A: Non-local Feynman rules}
\\[4pt]

\noindent In this appendix we list the non-local contributions to the various 
two-point and three-point interactions (see Ref.~\cite{Beenakker:1999hi} for 
more details). We start off with the two-point interactions. 
In order to calculate the non-local propagators we add the local 
gauge-fixing lagrangian corresponding to the covariant $R_{\xi}$ gauge:
\begin{eqnarray}
  \LL_{R_{\xi}}(x) &=& - \,\frac{1}{2}\,\Biggl\{
        \frac{1}{\xi_{\gamma}}\,\Bigl[\partial^{\mu} A_{\mu}(x)\Bigr]^{2}
        + \frac{1}{\xi_{_Z}}\,\Bigl[\partial^{\mu} Z_{\mu}(x) 
                                    - \xi_{_Z}M_{Z}\chi(x)\Bigr]^{2} 
        \nonumber \\
                   && \hspace*{-8ex}{}+\frac{2}{\xi_{_W}}\, 
        \Bigl[\partial^{\mu} W_{\mu}^{+}(x) - i \xi_{_W}M_{W}\phi^{+}(x)\Bigr]
        \Bigl[\partial^{\nu} W_{\nu}^{-}(x) + i \xi_{_W}M_{W}\phi^{-}(x)\Bigr]
                                                          \Biggr\}~,
\end{eqnarray}
where $\,\xi_{\gamma}\,$, $\,\xi_{_Z}\,$ and $\,\xi_{_W}\,$ are gauge 
parameters. Taking into account all local and non-local bilinear interactions 
we find the following (dressed) gauge-boson propagators after inversion: 
($V=\gamma,\,Z,\,W$)
\begin{eqnarray}
\label{prop}
  P^{VV}_{\mu\nu}(q,\xi_{_V}) &=& P^{VV}_{T,\,\mu\nu}(q) 
                                  + P^{VV}_{L,\,\mu\nu}(q,\xi_{_V})
                                  \nonumber \\[2mm]
                              &=& {}- i\,D_T^{VV}(q^2)\,
          \biggl( g_{\mu\nu} - \frac{q_{\mu}q_{\nu}}{q^2} \biggr)
                                  - i\,D_L^{VV}(q^2,\xi_{_V})\,
          \frac{q_{\mu}q_{\nu}}{q^2} \nonumber \\[1mm]
  P^{\gamma Z}_{\mu\nu}(q)    &=& P^{Z\gamma}_{\mu\nu}(q)
          \ =\ P^{\gamma Z}_{T,\,\mu\nu}(q) \ =\ {}- i\,D_T^{\gamma Z}(q^2)\,
          \biggl( g_{\mu\nu} - \frac{q_{\mu}q_{\nu}}{q^2} \biggr)~. 
\end{eqnarray}
Writing $q^2=s$, the gauge-boson propagator functions $D_T$ and $D_L$ are 
given by
\begin{eqnarray}
  D_T^{WW}(s)           &=& 
     \biggl\{ s\,\Bigl[1+\sigtwo(s)\Bigr] - M_W^2\Bigl[1+\sigfive(s)\Bigr]
     \biggr\}^{-1} \nonumber \\[1mm]
  D_L^{WW}(s,\xi_{_W})  &=& \xi_{_W}\, 
     \frac{s\,\Bigl[1+\sigfive(s)\Bigr] - \xi_{_W}\MW^2}
          {\Bigl[1+\sigfive(s)\Bigr] (s-\xi_{_W}\MW^2)^2} \nonumber \\[1mm]
  D_T^{\gamma\gamma}(s) &=&
     s\,\biggl\{ 1 + \sw^2\Bigl[\sigone(s)-2\sigthree(s)\Bigr] 
                   + \cw^2\Bigl[\sigtwo(s)+\sigfour(s)\Bigr] \biggr\}/D(s)
     \nonumber \\[1mm]
                          & &
     {}- \MZ^2\Bigl[1+\sigfive(s)+\sigsix(s)\Bigr]/D(s) \nonumber \\[1mm]
  D_L^{\gamma\gamma}(s,\xi_{\gamma}) &=& \xi_{\gamma}\, \frac{1}{s} 
     \nonumber \\[1mm]
  D_T^{ZZ}(s)           &=& 
     s\,\biggl\{1 + \sw^2\Bigl[\sigtwo(s)+2\sigthree(s)+\sigfour(s)\Bigr]
                  + \cw^2\sigone(s) \biggr\}/D(s) \nonumber 
\end{eqnarray}
\begin{eqnarray}
  D_L^{ZZ}(s,\xi_{_Z})           &=& \xi_{_Z}\,
     \frac{s\,\Bigl[1+\sigfive(s)+\sigsix(s)\Bigr] - \xi_{_Z}\MZ^2}
          {\Bigl[1+\sigfive(s)+\sigsix(s)\Bigr] (s-\xi_{_Z}\MZ^2)^2}
     \nonumber \\[1mm]
  D_T^{\gamma Z}(s)     &=& 
     {}- s\,\sw\cw\biggl[ \sigone(s)-\sigtwo(s)-\sigthree(s)
                         -\sigfour(s)+\frac{\sw^2}{\cw^2}\,\sigthree(s) 
                  \biggr]/D(s)~, \nonumber\\
\label{propagators}
\end{eqnarray}
with
\begin{eqnarray}
\label{D(s)}
  D(s) &=& s^2\,\biggl\{ \Bigl[1+\sigone(s)\Bigr]\,
                         \Bigl[1+\sigtwo(s)+\sigfour(s)\Bigr]
                         - \frac{\sw^2}{\cw^2}\,\Bigl[\sigthree(s)\Bigr]^2 
                \biggr\} \nonumber \\[1mm]
       & & \hspace*{-6ex}
           {}- s\,\MZ^2\Bigl[1+\sigfive(s)+\sigsix(s)\Bigr]
           \biggl\{ 1 + \sw^2\Bigl[\sigtwo(s)+2\sigthree(s)+\sigfour(s)\Bigr] 
                    + \cw^2\sigone(s) \biggr\}~. \nonumber\\
\end{eqnarray}
Note that $\,D_T^{WW}(0) = D_L^{WW}(0)\,$ and $\,D_T^{ZZ}(0) = D_L^{ZZ}(0)\,$ 
as a result of analyticity requirements. 

The dressed propagators involving the would-be Goldstone bosons are relatively
simple:
\begin{eqnarray}
  P^{W^{\pm}\phi^{\mp}}_{\mu}(q,\xi_{_W}) &=& 
     P^{\phi^{\pm}W^{\mp}}_{\mu}(q,\xi_{_W}) 
     \ = \ {}\pm i q_{\mu}\,\frac{\xi_{_W}\MW\sigfive(s)}
         {\Bigl[1+\sigfive(s)\Bigr] (s-\xi_{_W}\MW^2)^2}
     \nonumber \\[1mm]
                                          &\equiv& 
     q_{\mu}\,P^{W^{\pm}\phi^{\mp}}(s,\xi_{_W})
     \nonumber \\[3mm]
  P^{\phi\phi}(s,\xi_{_W}) &=& i\,
     \frac{s - \xi_{_W}\MW^2\Bigl[1+\sigfive(s)\Bigr]}
          {\Bigl[1+\sigfive(s)\Bigr] (s-\xi_{_W}\MW^2)^2}  
     \nonumber \\[1mm]
  P^{Z\chi}_{\mu}(q,\xi_{_Z}) &=& {}- P^{\chi Z}_{\mu}(q,\xi_{_Z})
     \ = \ {}- q_{\mu}\,\frac{\xi_{_Z}\MZ\Bigl[\sigfive(s)+\sigsix(s)\Bigr]}
         {\Bigl[1+\sigfive(s)+\sigsix(s)\Bigr] (s-\xi_{_Z}\MZ^2)^2}
     \nonumber \\[1mm]
                                          &\equiv& 
     q_{\mu}\,P^{Z\chi}(s,\xi_{_Z}) \nonumber \\[3mm]
  P^{\chi\chi}(s,\xi_{_Z}) &=& i\,
     \frac{s - \xi_{_Z}\MZ^2\Bigl[1+\sigfive(s)+\sigsix(s)\Bigr]}
          {\Bigl[1+\sigfive(s)+\sigsix(s)\Bigr] (s-\xi_{_Z}\MZ^2)^2}~.
\end{eqnarray}


Now we come to the three-point vertices. For a compact notation we first 
introduce the following three tensor structures:
\begin{align}
  T^{\mu_1\mu_2}(q_1,q_2) & = (q_1 q_2) g^{\mu_1\mu_2} - q_1^{\mu_2}q_2^{\mu_1}
                              \nn\\[4mm]
  A^{\mu_1,\mu_2\mu_3}(q) & = g^{\mu_1\mu_2}q^{\mu_3} - g^{\mu_1\mu_3}q^{\mu_2}
                              \nn\\[1mm]
  \A_5^{\mu}(q_1,q_2) & = \frac{\sigfive(q_2^2)-\sigfive(q_1^2)}{q_2^2-q_1^2}\,
                          (q_2-q_1)^{\mu}~.
\end{align}
In terms of these tensor structures the non-local three-point interactions read
\\[12pt]
\begin{eqnarray}
\hspace*{2ex}
\begin{picture}(80,30)(0,-2)
         \Text(-8,-15)[lb]{$q_1 \rightarrow $}
         \Photon(-8,0)(30,0){2}{5}
         \Text(-10,8)[lb]{$V_1,\mu_1$}
         \GCirc(30,0){3}{0}
         \Photon(30,0)(60, 30){2}{5}
         \Text(45,35)[lb]{$V_2,\mu_2$}
         \Text(62, 20)[lb]{$q_2$}
         \Text(49,9)[lb]{$ \swarrow $} 
         \Photon(30,0)(60,-30){2}{5}
         \Text(45,-47)[lb]{$V_3,\mu_3$}
         \Text(62,-27)[lb]{$q_3$}
         \Text(49,-20)[lb]{$ \nwarrow $}
\end{picture} \hspace*{-3ex} 
&:& ig_2 \Bigg\{ A_2 \sumperm \epsilon^{jkl}\,\sigtwo(q_j^2)\, 
                 \biggl[ \,\frac{(2q_j+q_l)^{\mu_l}}{(q_j+q_l)^2-q_j^2}\, 
                         T^{\mu_j\mu_k}(q_j,q_k) \nn\\
&& \hphantom{ig_2a} {}+ \frac{1}{2}\,A^{\mu_j,\mu_k\mu_l}(q_j)\, \biggr]
                    + A^{\mu_1,\mu_2\mu_3}(q_1)\,\Bigl[ A_{31}\sigthree(q_1^2) 
                    + A_{41}\sigfour(q_1^2) \Bigr] \nn\\
&& \hphantom{ig_2a} {}+ \frac{\MW^2}{2\,\cw}\,\sumperm A_{jkl}\,\,
                        g^{\mu_j\mu_k}\,\A_5^{\mu_l}(q_j,q_k) \Bigg\}~.
\label{TGC_nonlocal}
\end{eqnarray}
The summation includes all possible permutations $(j,k,l)$ of $(1,2,3)$ and
the various couplings are given by
\begin{equation}  
\begin{array}{|c|c|c|c|c|}\cline{1-5}
V_1 V_2 V_3   & A_2  & A_{31}    & A_{41} 
              & \mbox{non-zero coefficients }A_{jkl} \\ \hline
\gamma W^+W^- & \sw  & \sw       & \sw   
              & A_{231} = 2\,\cw\sw                  \\
Z W^+W^-      & -\cw & \sw^2/\cw & -\cw   
              & A_{123} = - A_{132} = -1\ , \  
                A_{231} = \sw^2-\cw^2                \\[-5mm] 
              & & & & \\ \cline{1-5} 
\end{array}  \nn
\end{equation} 
 
\vspace{1cm}

\begin{eqnarray}
\begin{picture}(80,30)(0,-2)
         \Text(0,10)[lb]{$S$}
         \DashLine(0,0)(30,0){5}
         \Text(0,-12)[lb]{$q \rightarrow $}
         \GCirc(30,0){3}{0}
         \Photon(30,0)(60, 30){2}{5}
         \Text(45,35)[lb]{$V_1,\mu_1$}
         \Text(62, 19)[lb]{$q_1$}
         \Text(48,9)[lb]{$\swarrow $}
         \Photon(30,0)(60,-30){2}{5}
         \Text(45,-45)[lb]{$V_2,\mu_2$}
         \Text(61,-26)[lb]{$q_2$}
         \Text(48,-19)[lb]{$\nwarrow $}
\end{picture} \hspace*{-3ex}
&:& \frac{ig_2}{\MW}\Biggl\{ T^{\mu_1\mu_2}(q_1,q_2)\,\biggl[
    \,\frac{\sw}{\cw}\,\Bigl[ C_{31}\sigthree(q_1^2) 
                            + C_{32}\sigthree(q_2^2) \Bigr]
    + C_{41}\sigfour(q_1^2) \nn\\ 
&&  \hphantom{\frac{ig_2}{\MW}a}{}+ C_{42}\sigfour(q_2^2)\, \biggr]
    - \frac{\MW^2}{2\,\cw}\,\sumperm\,
    \biggl[ \,C_{5jk}\,\Bigl[ q^{\mu_j}\,\A_5^{\mu_{k}}(q_j,q)\nn\\
&&  \hphantom{\frac{ig_2}{\MW}a}{}+ g^{\mu_j\mu_k}\,\sigfive(q_j^2) \Bigr]
    + 2\,g^{\mu_j\mu_k}\,C_{6jk}\,\sigsix(q_j^2)\, \biggr] \Biggr\}~. 
\end{eqnarray}
The summation includes all possible permutations $(j,k)$ of $(1,2)$ and
the various couplings are given by
\begin{equation} 
\begin{array}{|c|c|c|c|c|c|c|c|c|} \cline{1-9}
SV_1V_2                  & \!C_{31}\!      & \!C_{32}\!  & \!C_{41}\!  
                         & \!C_{42}\!      & \!C_{512}\! & \!C_{521}\!     
                         & \!C_{612}\!     & \!C_{621}\! \\ \hline
HZZ                      & \!-\sw\cw\!     & \!-\sw\cw\! & \!\cw^2\!   
                         & \!\cw^2\!       & \!-1/\cw\!  & \!-1/\cw\!      
                         & \!-1/\cw\!      & \!-1/\cw\!  \\ 
HZ\gamma                 & \!\sw^2\!       & \!-\cw^2\!  & \!-\sw\cw\! 
                         & \!-\sw\cw\!     & \!0\!       & \!0\!           
                         & \!0\!           & \!0\!       \\
H\gamma\gamma            & \!\sw\cw\!      & \!\sw\cw\!  & \!\sw^2\!   
                         & \!\sw^2\!       & \!0\!       & \!0\!           
                         & \!0\!           & \!0\!       \\
H W^+ W^-                & \!0\!           & \!0\!       & \!0\!       
                         & \!0\!           & \!-\,\cw\!  & \!-\,\cw\!      
                         & \!0\!           & \!0\!       \\
\chi W^+ W^-             & \!0\!           & \!0\!       & \!0\!       
                         & \!0\!           & \!i\,\cw\!  & \!-i\,\cw\!     
                         & \!0\!           & \!0\!       \\ 
\phi^{\mp}ZW^{\pm}       & \!\sw\!         & \!0\!       & \!-\cw\!    
                         & \!0\!           & \!1\!       & \!\sw^2-\cw^2\! 
                         & \!1\!           & \!0\!       \\
\phi^{\mp}\gamma W^{\pm} & \!\cw\!         & \!0\!       & \!\sw\!     
                         & \!0\!           & \!0\!       & \!2\sw\cw\!     
                         & \!0\!           & \!0\!       \\[-5mm]
                         & & & & & & & & \\ \cline{1-9}
\end{array} \nn
\end{equation} 

\begin{eqnarray} 
\hspace*{1ex}
\begin{picture}(75,55)(0,-2)
   \Photon(-8,0)(30,0){2}{5}
   \Text(-10,8)[lb]{$V_1,\mu_1$}
   \Text(-8,-15)[lb]{$q_1 \rightarrow $}
   \GCirc(30,0){3}{0}
   \DashLine(30,0)(60, 30){5}
   \Text(55, 35)[lb]{$S_2$}
   \Text(62, 20)[lb]{$q_2$}
   \Text(49,9)[lb]{$ \swarrow $}
   \DashLine(30,0)(60,-30){5}
   \Text(55,-45)[lb]{$S_3$}
   \Text(62,-27)[lb]{$q_3$}
   \Text(49,-20)[lb]{$ \nwarrow $}
\end{picture} \hspace*{-1ex}
&:& {} -\frac{i\,e}{2\,\cw\sw}\,
       \bigg\{ E_1\Bigl[ (q_2 q_3)\,\A_5^{\mu_1}(q_2,q_3)
                          + q_2^{\mu_1}\,\sigfive(q_2^2)
                          - q_3^{\mu_1}\,\sigfive(q_3^2) \Bigr] \nn\\ 
&& \hphantom{-\frac{i\,e}{2\,\cw\sw}Aa}
               {}+ E_{23}\,(q_2-q_3)^{\mu_1}\,\sigsix(q_1^2)
               + E_3\,q_3^{\mu_1}\,\sigsix(q_3^2) \bigg\}~.
\end{eqnarray}
The various couplings are given by
\begin{equation}
  \begin{array}{|c|c|c|c|} \cline{1-4} 
    V_1S_2S_3             & E_1         & E_{23} & E_3        \\ \hline
    Z\phi^+\phi^-         & \sw^2-\cw^2 & 1      & 0          \\
    \gamma\phi^+\phi^-    & 2\,\cw\sw   & 0      & 0          \\
    Z H\chi               & -\,i        & -\,i   & 2\,i       \\
    W^{\pm} H\phi^{\mp}   & \mp\,\cw    & 0      & 0          \\
    W^{\pm}\phi^{\mp}\chi & i\,\cw      & 0      & -2\,i\,\cw \\ \cline{1-4}
\end{array} \nn
\end{equation}

\vspace{4 mm}

\begin{eqnarray}
\begin{picture}(80,30)(0,-2)
         \Text(-5,10)[lb]{$S_1$}
         \Text(-8,-15)[lb]{$q_1 \rightarrow $}
         \DashLine(0,0)(30,0){5}
         \GCirc(30,0){3}{0}
         \DashLine(30,0)(60, 30){5}
         \Text(55, 35)[lb]{$S_2$}
         \Text(62, 20)[lb]{$q_2$}
         \Text(49,9)[lb]{$ \swarrow $}
         \DashLine(30,0)(60,-30){5}
         \Text(55,-45)[lb]{$S_3$}
         \Text(62,-27)[lb]{$q_3$}
         \Text(49,-20)[lb]{$ \nwarrow $}
\end{picture}
&:& {}-\frac{i}{v}\,\sumperm (q_j q_l)\,F_{jkl}\,\sigsix(q_j^2)~.
\end{eqnarray}
{}\vspace*{4ex}\\
The summation includes all possible permutations $(j,k,l)$ of $(1,2,3)$ and
the various couplings are given by
\begin{equation}
\begin{array}{|l|c|}
    \hline
    S_1S_2S_3        & \mbox{non-zero coefficients }F_{jkl}                \\
    \hline           &  \\[-4.5mm]
    HHH              & F_{jkl}=1 \ \ 
                       \mbox{ for all permutations of (1,2,3)}             \\
    H\chi\chi        & F_{123}=F_{132}=-F_{231}=-F_{321}=F_{213}=F_{312}=1 \\
    H\phi^+\phi^-    & F_{123}=F_{132}=1                                   \\
    \chi\phi^+\phi^- & F_{123}=-F_{132}=i                                  \\
    \hline
  \end{array} \nn
\end{equation}

\vspace*{18pt}
\noindent {\large\bf Appendix B: Two-point functions in the \FL scheme}
\\[4pt]

\noindent In this appendix we list the various unrenormalized fermion-loop 
self-energies in the SM, which will be needed for determining the six 
non-local coefficients.

The gauge-boson self-energies can be written as: 
\begin{equation}
  \begin{picture}(80,30)(0,-2)
    \Photon(0,0)(80,0){2}{10}
    \ArrowLine(25,-12)(26,-12)
    \Line(15,-12)(25,-12)
    \ArrowLine(55,-12)(54,-12)
    \Line(65,-12)(55,-12)
    \Text(0,8)[lb]{$V_1,\mu$}
    \Text(0,-8)[lt]{$q$}
    \Text(80,8)[rb]{$V_2,\nu$}
    \Text(74,-8)[lt]{$-\,q$}
    \GCirc(40,0){10}{1}
  \end{picture} 
  \hspace*{3ex}
  :\ i\,\Sigma_{\mu\nu}^{V_1V_2}(q) 
  \ =\  
  {}-i\,\Sigma_T^{V_1V_2}(q^2)\,\Bigl( g_{\mu\nu} 
                                     - \frac{q_{\mu}q_{\nu}}{q^2} \Bigr)
  - i\,\Sigma_L^{V_1V_2}(q^2)\,\frac{q_{\mu}q_{\nu}}{q^2}~,
  \vspace*{2ex}
\end{equation}
where $\Sigma_T$ and $\Sigma_L$ are the transverse and longitudinal
gauge-boson self-energies, respectively. Writing $q^2=s$, we find for the 
transverse self-energies
\begin{eqnarray}
\label{pifgamma}
  \Sigma_T^{\gamma\gamma}(s) &=& \frac{\alpha}{3\pi}\,\sumf Q_f^2 
          \Biggl\{ s\,\biggl[ B_0(s,0,0) - \frac{1}{3} \biggr]
        + s\,\Bigl[ B_0(s,m_f,m_f) - B_0(s,0,0) \Bigr] \nonumber \\[1mm]
                             & & \hspace*{-4ex}
        {}+ 2m_f^2\,\Bigl[ B_0(s,m_f,m_f) - B_0(0,m_f,m_f) \Bigr] \Biggr\}
                           \ \equiv \ s\,\sumf  Q_f^2\,\pifs 
\end{eqnarray}
and
\begin{eqnarray}
\label{SigmaT}
 \Sigma_T^{\gamma Z}(s)     &=& -\,s\,\sumf   
          \Biggl[ \frac{|Q_f|}{4\sw\cw} - \frac{\sw}{\cw}\,Q_f^2 \Biggr] \pifs
          \nonumber \\[1mm]
  \Sigma_T^{ZZ}(s)           &=& s\,\sumf   
          \Biggl[ \frac{(\cw^2-\sw^2)|Q_f|}{4\sw^2\cw^2}
                + \frac{\sw^2}{\cw^2}\,Q_f^2 \Biggr] \pifs 
        + \frac{T_{_Z}(s)}{\cw^2} \nonumber \\[1mm]
  \Sigma_T^{WW}(s)           &=& s\,\sumf \frac{|Q_f|}{4\sw^2}\,\pifs
        + T_{_W}(s)~.
\end{eqnarray}
The scalar two-point functions $B_0$ are defined in the usual 
way~\cite{Denner:kt} and 
\begin{eqnarray}
  T_{_Z}(s) &=& 2\MW^2 T  
        - \frac{\alpha}{8\pi\sw^2}\,\sumf m_f^2\,B_0(s,m_f,m_f)
        - s\,\sumf \frac{2\,|Q_f|\!-\!1}{8\sw^2}\,\pifs 
          \nonumber \\[1mm]
  T_{_W}(s) &=& T_{_Z}(s) + \frac{\alpha}{24\pi\sw^2}\,\sumf \biggl\{
          (s-m_f^2) \Bigl[ B_0(s,m_{f'},m_f) - B_0(s,m_f,m_f) \Bigr]
          \nonumber \\[1mm]
            & &
        {}- \frac{m_f^2(m_f^2-m_{f'}^2)}{s} \Bigl[ B_0(s,m_{f'},m_f) 
                                                 - B_0(0,m_{f'},m_f) \Bigr] 
                                                             \biggr\}~.
\end{eqnarray}
The constant $\,T\,$ represents the universal tadpole contribution, which does 
not need to be specified in view of the fact that we will perform tadpole 
renormalization anyhow (see later). 
In a similar way the longitudinal gauge-boson self-energies are given by   
\begin{eqnarray}
\label{SigmaL} 
  \Sigma_L^{\gamma\gamma}(s) &=& 0 \nonumber \\[3mm]
  \Sigma_L^{\gamma Z}(s)     &=& 0 \nonumber \\[3mm]
  \Sigma_L^{ZZ}(s)           &=& 2\MZ^2\,T 
        - \frac{\alpha}{8\pi\sw^2\cw^2}\,\sumf m_f^2\,B_0(s,m_f,m_f) 
          \nonumber \\[1mm] 
  \Sigma_L^{WW}(s)           &=& 2\MW^2 T  
        + \frac{\alpha}{8\pi\sw^2}\,\sumf m_f^2\,
          \biggl\{ {}- B_0(s,m_{f'},m_f) \nonumber \\[1mm]
                             & &
          {}+ \frac{m_f^2-m_{f'}^2}{s} \Bigl[ B_0(s,m_{f'},m_f) 
                                            - B_0(0,m_{f'},m_f) \Bigr] 
          \biggr\}~.
\end{eqnarray}

In the longitudinal/scalar sector there are a few more 
self-energies to be considered:
\begin{eqnarray}
  \begin{picture}(80,30)(0,-2)
    \Photon(0,0)(40,0){2}{5}
    \DashLine(40,0)(80,0){5}
    \ArrowLine(25,-12)(26,-12)
    \Line(15,-12)(25,-12)
    \ArrowLine(55,-12)(54,-12)
    \Line(65,-12)(55,-12)
    \Text(0,8)[lb]{$V_1,\mu$}
    \Text(0,-8)[lt]{$q$}
    \Text(80,8)[rb]{$S_2$}
    \Text(74,-8)[lt]{$-\,q$}
    \GCirc(40,0){10}{1}
  \end{picture} 
  \hspace*{5ex}
  &:& i\,\Sigma_{\mu}^{V_1S_2}(q) 
  \ =\  
  i\,q_{\mu}\,\Sigma^{V_1S_2}(q^2)
  \\[3ex]
  \begin{picture}(80,30)(0,-2)
    \DashLine(0,0)(80,0){5}
    \ArrowLine(25,-12)(26,-12)
    \Line(15,-12)(25,-12)
    \ArrowLine(55,-12)(54,-12)
    \Line(65,-12)(55,-12)
    \Text(0,8)[lb]{$S_1$}
    \Text(0,-8)[lt]{$q$}
    \Text(80,8)[rb]{$S_2$}
    \Text(74,-8)[lt]{$-\,q$}
    \GCirc(40,0){10}{1}
  \end{picture} 
  \hspace*{5ex}
  &:& i\,\Sigma^{S_1S_2}(q^2)~. \\
  \nonumber \vspace*{1ex}
\end{eqnarray}
These self-energy functions can be expressed in terms of the longitudinal 
gauge-boson self-energies given above:
\begin{eqnarray}
  \Sigma^{Z\chi}(s) &=& {}- \Sigma^{\chi Z}(s) 
          \ =\ {}- \frac{i}{\MZ}\,\biggl[ \Sigma_L^{ZZ}(s) - \MZ^2\,T \biggr]
          \nonumber \\[1mm] 
  \Sigma^{\chi\chi}(s) &=& {}- \frac{s}{\MZ^2}\,\biggl[ \Sigma_L^{ZZ}(s) 
          - 2\MZ^2\,T \biggr] \nonumber \\[1mm]
  \Sigma^{W^{\pm}\phi^{\mp}}(s) &=& \Sigma^{\phi^{\pm}W^{\mp}}(s)  
          \ =\ \mp\,\frac{1}{\MW}\,\biggl[ \Sigma_L^{WW}(s) - \MW^2 T \biggr] 
          \nonumber \\[1mm]
  \Sigma^{\phi\phi}(s) &=& {}- \frac{s}{\MW^2}\,\biggl[ \Sigma_L^{WW}(s) 
          - 2\MW^2 T \biggr]~.  
\end{eqnarray}
For completeness we also give the fermion-loop self-energy of the physical 
Higgs boson:
\begin{eqnarray}
  \Sigma^{HH}(s) &=& 3M_{_H}^2\,T - \frac{1}{8\pi^2 v^2}\,\sumf m_f^2\,
                     \biggl\{ (4m_f^2-s)B_0(s,m_f,m_f) \nn\\[1mm]
                 & & {}+ 2m_f^2 \Bigl[ B_0(0,m_f,m_f) + 1 \Bigr] \biggr\}~.
\end{eqnarray}

The above-given longitudinal/scalar self-energies satisfy the following Ward
Identities:
\begin{eqnarray}
  \Sigma_L^{ZZ}(s) - 2i\MZ\Sigma^{Z\chi}(s) 
  - \frac{\MZ^2}{s}\,\Sigma^{\chi\chi}(s) &=& 0 \\[1mm]
  \Sigma_L^{WW}(s) \pm 2\MW\Sigma^{W^{\pm}\phi^{\mp}}(s)
  - \frac{\MW^2}{s}\,\Sigma^{\phi\phi}(s) &=& 0~.
\end{eqnarray}

As a next step we perform tadpole renormalization. This involves shifting the
bare vacuum in such a way that at one-loop level it coincides with the true 
vacuum of the Higgs potential. Or in other words, the one-point counterterm 
generated by the finite shift of the bare vacuum $v$ completely compensates
the tadpole self-energy terms $\propto T$. This is equivalent to the following 
effective procedure: keep the bare vacuum as it is, but remove the terms 
$\propto T$ from the $WW,\,ZZ,\,W\phi,\,Z\chi$, and $HH$ self-energies. 
The $\phi\phi$ and $\chi\chi$ self-energies receive both one-point and 
two-point counterterms, which exactly cancel each other. So, the tadpole 
contributions to these self-energies should be kept and are therefore merged 
with the rest of the fermion-loop corrections. This is a trivial exercise,
since the $\phi\phi$ and $\chi\chi$ self-energies have an internal 
cancellation of all terms $\propto T$ (see the expressions above).

\vspace*{18pt}
\noindent {\large\bf Appendix C: Renormalization conditions}
\\[4pt]

\noindent An essential ingredient of the matching procedure is the 
renormalization of the non-local coefficients, which takes the form of matching
the non-local matrix elements and cross sections with explicit experimental 
observables. Various options are open, each with their own merits. Let us go 
through the most popular renormalization/matching conditions, bearing in mind 
that we only have to fix the running couplings $\,v^2(s),\,1/g_2^2(s)\,$ and 
$\,1/e^2(s)$.

\paragraph{\bf Muon decay:}
One of the often applied matching conditions 
is based on the charged-current
muon-decay process $\mu^- \to \nu_{\mu}e^-\bar{\nu}_e$. In the unitary gauge 
we obtain the matrix element
\begin{equation}
  \M_1 = {}-\frac{g_2^2}{2}\,P^{WW}_{\rho\sigma}(q_{_W},\xi_{_W}\to\infty)\,
         [\bar{u}_{\nu_{\mu}}(p_{\nu_{\mu}}) \gamma^{\rho} \omega_- 
          u_{\mu}(p_{\mu})]\,
         [\bar{u}_e(p_e) \gamma^{\sigma} \omega_- v_{\nu_e}(p_{\bar{\nu}_e})]~,
\end{equation}
with $\,q_{_W}=p_e+p_{\bar{\nu}_e}\,$ and $\,\omega_{\pm}=(1\pm\gamma_5)/2\,$. 
Upon neglecting $m_e$ and $m_{\mu}$ with respect to $\MW$, the expression 
simplifies to
\begin{equation}
  \M_1 = \frac{i}{\V_1(q_{_W}^2)}\,
         [\bar{u}_{\nu_{\mu}}(p_{\nu_{\mu}}) \gamma^{\rho} \omega_- 
          u_{\mu}(p_{\mu})]\,
         [\bar{u}_e(p_e) \gamma_{\rho}\,\omega_- v_{\nu_e}(p_{\bar{\nu}_e})]
\end{equation}
in terms of the inverse amplitude
\begin{equation}
  \V_1(q_{_W}^2) = 2\,q_{_W}^2\,\biggl[ \frac{1}{g_2^2(q_{_W}^2)} 
                   - \Sthree(q_{_W}^2) - \Sfour(q_{_W}^2) \biggr]
                   - \frac{1}{2}\,v^2(q_{_W}^2)~.
\end{equation}
This final step was obtained with the help of Eqs.(\ref{running2}), 
(\ref{unitary_gauge}) and (\ref{calW/Z}).
In the muon-decay process we can go one step further, since  
$q_{_W}^2 = {\cal O}(m_{\mu}^2) \ll \MW^2$. In that case the (low-energy) 
inverse amplitude reads $\V_1(0) = -\,v^2(0)/2$.
The actual matching condition links this inverse amplitude to the 
experimentally-determined coefficient of the effective (low-energy) 
charged-current $V\!-\!A$ lagrangian
\begin{equation} 
  \LL_{\sss{eff}} = {}-2\sqrt{2}\,G_F\,
            [\bar{\psi}_{\nu_{\mu}} \gamma^{\rho} \omega_- \psi_{\mu}]\,
            [\bar{\psi}_e \gamma_{\rho}\,\omega_- \psi_{\nu_e}] + \cdots
\end{equation}
In this way we obtain
\begin{equation}
\label{v-matching}
  v^2(0) \ =\ v^2 + \Sfive(0) \ = \ \frac{1}{\sqrt{2}\,G_F} 
  \ \Rightarrow \ 
  v^2(s) = \frac{1}{\sqrt{2}\,G_F} + \Sfive(s) - \Sfive(0)~.
\end{equation}

\paragraph{\boldmath The $W$-boson mass:}
A second, optional matching condition involves the mass of the $W$ bosons. 
For the definition we can again make use of the inverse amplitude 
$\V_1(q_{_W}^2)$. The most commonly used procedure is the on-shell
condition
\begin{equation} 
\label{mW-matching}
  \real\Bigl[ \V_1(m_{_W}^2) \Bigr] = 0
  \ \Rightarrow \ 
  \frac{1}{g_2^2} \ = \ \frac{\real v^2(m_{_W}^2)}{4m_{_W}^2} 
                        - \real \Stwo(m_{_W}^2)~,
\end{equation}  
where $m_{_W}$ is the experimentally-determined $W$-boson mass (based on an
on-shell analysis). This results in
\begin{equation} 
  \frac{1}{g_2^2(s)} = \frac{\real v^2(m_{_W}^2)}{4m_{_W}^2} 
           - \real \Stwo(m_{_W}^2) + \Stwo(s) + \Sthree(s) + \Sfour(s)~.
\end{equation} 

The on-shell procedure breaks down if one includes two-loop corrections
\cite{Sirlin:fd}, therefore it is sometimes better to use the complex $W$-boson
pole $\muW$ in the matching procedure:
\begin{equation} 
\label{muW-matching}
  \V_1(\muW) = 0
  \ \Rightarrow \ 
  \frac{1}{g_2^2} \ = \ \frac{v^2(\muW)}{4\muW} - \Stwo(\muW)~.
\end{equation} 
The real part of this complex pole can now be identified with  
the experimentally-determined $W$-boson mass, provided the same complex 
procedure is adopted in the data analysis.

\paragraph{\boldmath The $Z$-boson mass:}
A very precisely known experimental observable is the $Z$-boson mass, so it is
a natural candidate for performing the matching. A defining process for the 
$Z$-boson mass is the reaction $\nu_e\bar{\nu}_e \to \nu_{\mu}\bar{\nu}_{\mu}$,
which leads in the unitary gauge to the following (inverse) amplitude:
\begin{eqnarray}
  \M_2 &=& \frac{i}{{\cal V}_2(q_{_Z}^2)}\,
           [\bar{u}_{\nu_{\mu}}(p_{\nu_{\mu}}) \gamma^{\rho} \omega_- 
            v_{\nu_{\mu}}(p_{\bar{\nu}_{\mu}})]\,
           [\bar{v}_{\nu_e}(p_{\bar{\nu}_e}) \gamma_{\rho}\,\omega_- 
            u_{\nu_e}(p_{\nu_e})] 
           \nonumber \\[1mm]
  \V_2(q_{_Z}^2) &=& 4\,q_{_Z}^2\,\biggl[ 
           \frac{\cw^2(q_{_Z}^2)}{g_2^2(q_{_Z}^2)} - \Sthree(q_{_Z}^2) \biggr]
           - v^2(q_{_Z}^2) - \Ssix(q_{_Z}^2)~,
\end{eqnarray}
with $\,q_{_Z}=p_{\nu_e}+p_{\bar{\nu}_e}$. This expression was again obtained 
with the help of Eqs.(\ref{running2}), (\ref{unitary_gauge}) and 
(\ref{calW/Z}). Using Eqs.(\ref{running1}) and (\ref{running2}), the 
experimentally-determined value of the $Z$-boson mass in the
on-shell approach ($m_{_Z}$) and the on-shell condition 
$\real \V_2(m_{_Z}^2) = 0$, one arrives at the following quadratic equation:
\begin{eqnarray} 
  0 &=& \frac{\real A}{g_2^4} + \frac{\real B}{g_2^2} + \real C  
        \nonumber \\[2mm]
  A &=& e^2(m_{_Z}^2) \nonumber \\[3mm]
  B &=& 2\,e^2(m_{_Z}^2) \Bigl[ \Stwo(m_{_Z}^2) + \Sthree(m_{_Z}^2) 
                                + \Sfour(m_{_Z}^2) \Bigr]  - 1
        \nonumber \\[1mm]
  C &=& \frac{v^2(m_{_Z}^2) + \Ssix(m_{_Z}^2)}{4 m_{_Z}^2}
        + e^2(m_{_Z}^2) \Bigl[ \Stwo(m_{_Z}^2) + \Sthree(m_{_Z}^2) 
                               + \Sfour(m_{_Z}^2) \Bigr]^2
        \nonumber \\[1mm] 
    & & {}- \Stwo(m_{_Z}^2) - \Sfour(m_{_Z}^2)~. \hphantom{AA} 
\end{eqnarray}
The relevant solution is given by 
\begin{equation}
\label{mZ-matching}
\frac{1}{g_2^2} = \frac{1}{2\,\real A}\,\Bigl\{\, -\,\real B 
                  - \sqrt{(\real B)^2-4\,\real A\,\real C}\,\, \Bigr\}~.
\end{equation}
For a matching procedure based on the complex $Z$-boson pole, $\muZ$, one 
merely has to replace $\,\real A,\,\real B,\,\real C\,$ and $\,m_{_Z}^2$ 
by $\,A,\,B,\,C\,$ and $\muZ$.

\paragraph{\bf The electromagnetic coupling:}
The matching condition for the electromagnetic coupling has to be addressed 
with care, in view of its far-reaching consequences for the low- and 
high-scale behaviour of the cross sections. The complication is caused by the 
hadronic part of the photonic vacuum polarization, which is sensitive to 
non-perturbative strong-interaction effects through the exchange of gluons with
low momentum transfer.    
 
We can either match at a high scale, like $q_{\gamma}^2=m_{_Z}^2$, or at a low
scale, like the Thomson limit $q_{\gamma}^2=0$. In the former case we have to
evolve the running coupling down to low scales in order to deal with phenomena
that involve nearly on-shell photons (cf.~single $W$ production). In the latter
case we have to evolve the running coupling up to high scales in order to
properly describe high-scale reactions. From this it should be clear that 
preferably we want to match the complete running of the electromagnetic 
coupling, instead of matching it in a single point. To this end we have to 
exploit the explicit parametric dependence of the non-local coefficients, 
i.e.~the dependence on the fermion masses $m_f$. It has no use fiddling around 
with the lepton masses, since these masses are experimentally well-known and 
the leptonic evolution of the electromagnetic coupling is free of ambiguities.
The same does not apply to the hadronic part of the photonic vacuum 
polarization, in view of the various light-quark bound states that contribute.
So, the light-quark masses are prime candidates for tuning the evolution of the
electromagnetic coupling.

We start with the electromagnetic coupling at the LEP1 $Z$ peak. Usually this
coupling is presented for five active quark flavours and without imaginary 
part, i.e.~$\real [\alpha^{(5)}(m_{_Z}^2)^{-1}]$. The relation between 
$\alpha^{(5)}(s)$ and $\alpha(s)$ is fixed by the requirement of top-quark 
decoupling at $s=0$: 
\begin{equation} 
  \frac{1}{\alpha(s)} 
  =
  \frac{1}{\alpha} + \sumf Q_f^2\,\frac{\pifs}{\alpha}
  =
  \frac{1}{\alpha^{(5)}(s)} 
  + N_{_C}^t Q_t^2\,\frac{\Pi_t^{\gamma}(s) - \Pi_t^{\gamma}(0)}{\alpha}~.
\end{equation} 
By fixing the value of 
$\,\real [\alpha^{(5)}(m_{_Z}^2)^{-1}] \equiv \alpha_Z^{-1}\,$ we obtain
\begin{equation}
\label{alphaZ-matching} 
  \frac{1}{e^2} = \frac{1}{4\pi\alpha}
                = \frac{1}{4\pi\alpha_{_Z}} 
                  - \sum_{f\ne t} N_{_C}^f Q_f^2\,\real\!\left[
                    \frac{\Pi_f^{\gamma}(m_z^2)}{e^2} \right]
                  - N_{_C}^t Q_t^2\,\frac{\Pi_t^{\gamma}(0)}{e^2}~.
\end{equation} 
The leptonic as well as top-quark contributions can now be calculated 
perturbatively. Subsequently we tune the evolution to lower scales by using 
a set of effective light-quark masses. In the standard procedure this set of 
light-quark masses represents a perturbative fit to the 
once-subtracted dispersion integral of the experimental observable
$R^{\gamma}(s) = \frac{\displaystyle 3s}{\displaystyle 4\pi\alpha^2(0)}\,
                 \sigma(e^+e^- \to \gamma^* \to \mbox{hadrons})$: 
\begin{equation} 
  \sum\limits_{f=q,\,f \neq t} N_{_C}^f Q_f^2\,
  \frac{\pifs - \Pi_f^{\gamma}(0)}{4\pi\alpha}
  =
  \frac{s}{12\pi^2}\,\int\limits_{4m_{\pi}^2}^{\infty} \mbox{d}s'
  \frac{R^{\gamma}(s')}{s'(s'-s-i\varepsilon)}~, 
\end{equation}   
which automatically covers all non-perturbative hadronic contributions.
Note that in this way the quality of $\alpha(0)$ is linked to the quality of 
the effective-mass parametrization at the matching scale $s=m_{_Z}^2$ and to
the quality of the perturbative calculation used in the fit.
This can be circumvented by adopting an alternative procedure, which uses 
the value of $\,\alpha(0) \equiv \alpha_0\,$ as additional matching condition, 
leading to
\begin{equation} 
\label{alpha0-matching}
  \frac{1}{e^2} 
  =
  \frac{1}{4\pi\alpha_0} - \sumf Q_f^2\,\frac{\Pi_f^{\gamma}(0)}{e^2}~.
\end{equation} 
This second matching condition can be implemented by tuning the effective 
$u$- and $d$-quark masses. \vspace*{2ex}

Based on the matching conditions discussed above, various matching procedures 
are possible. 
\paragraph{1) \underline{The LEP1 procedure} :}
this procedure combines the measured top-quark mass at the Tevatron with three
of the above-mentioned matching conditions, i.e.~the muon-decay condition and 
the two on-shell LEP1 conditions for the $Z$-boson mass and the 
electromagnetic coupling. So, with this procedure the input parameters are
$G_F,\,m_{_Z},\,\real[\alpha^{(5)}(m_{_Z}^2)^{-1}]\equiv\alpha_{_Z}^{-1},
\,m_t$, the lepton masses and the standard set of effective light-quark masses.
\paragraph{2) \underline{The LEP2 procedure} :}
this procedure adds the on-shell $W$-boson mass $m_{_W}$ to the matching 
conditions of the LEP1 procedure. Because $1/g_2^2$ is now matched twice, we 
end up with a consistency relation. The dominant ingredient in this relation 
is the top-quark contribution. Therefore we can determine the top-quark mass by
solving the consistency relation iteratively. Since we exclusively take into 
account fermion loops, the resulting top-quark mass will be an effective 
parameter that mimics bosonic loop effects. Its value will come out appreciably
lower than the experimental measurement (see Ref.~\cite{Beenakker:1996kn}).
\paragraph{3) \underline{Our procedure} :}
in the present analysis we have added the electromagnetic coupling in the 
Thomson limit, $\alpha(0) \equiv \alpha_0$, to the matching conditions of the 
LEP2 procedure. This results in an alternative set of effective light-quark 
masses, with adjusted values for the $u$- and $d$-quark masses. 
Moreover, we have replaced the on-shell conditions for the $W$- and $Z$-boson 
masses by their complex counterparts. So, we use Eqs.(\ref{v-matching}),
(\ref{muW-matching}), (\ref{alphaZ-matching}) and (\ref{alpha0-matching}), as
well as the complex version of \eqn{mZ-matching}. This system of equations is
solved iteratively, resulting in the determination of the complex gauge-boson
poles $\mu_{_W, _Z}$, the effective top-quark mass $m_t$ and the effective 
common light-quark mass $m_u=m_d$.

\vspace*{18pt}
\noindent {\large\bf\boldmath Appendix D: $e^+e^-\to W_T^+W_T^-\,$ and 
           high-energy unitarity}
\\[4pt]

\noindent In this appendix we have a closer look at the process of on-shell 
transverse $W$-pair production, $e^+(p_1) e^-(p_2) \to W_T^+(p_+) W_T^-(p_-)$, 
in the high-energy limit. Neglecting the electron mass, the momenta of the 
particles are defined as follows in the initial-state centre-of-mass frame:
\begin{equation}
\begin{array}{l l} 
p_1=\frac{\displaystyle\sqrt{s}}{\displaystyle 2}\,(1,0,0,1)\, ,  & 
p_+=\frac{\displaystyle\sqrt{s}}{\displaystyle 2}\,(1,0,\beta\sj,\beta\cj)
\\[1mm]
p_2=\frac{\displaystyle\sqrt{s}}{\displaystyle 2}\,(1,0,0,-1)\, , & 
p_-=\frac{\displaystyle\sqrt{s}}{\displaystyle 2}\,(1,0,-\beta\sj,-\beta\cj)~,
\\
\end{array} \nn
\end{equation}
where $\theta$ is the scattering angle and $\,\beta = \sqrt{1-4\MW^2/s}\,\,$ 
is the velocity of the $W$ bosons. In this frame the polarization vectors for 
transversely polarized $W$ bosons are given by
\begin{align}
\epsilon_{\pm}^{\mu}(p_+) & = \frac{1}{\sqrt{2}}\,(\,0,i,\mp\cj,\pm\sj) \nn\\
\epsilon_{\pm}^{\mu}(p_-) & = \frac{1}{\sqrt{2}}\,(\,0,-i,\mp\cj,\pm\sj)~, \nn
\end{align}
where the subscripts $\,\pm\,$ indicate the transverse helicities $\,\pm 1\,$ 
of the considered $W$ boson. Note that these polarization vectors are 
orthogonal with respect to both $p_+$ and $p_-$.
Finally, the left- and right-handed electron--positron currents can be written 
as
\begin{align} 
J^\mu_L&= \bar{v}_{e}(p_1) \gamma^{\mu} \omega_- u_{e}(p_2)
          = \sqrt{s}\,(0,-i,-1,0) \nn\\[1mm]
J^\mu_R&= \bar{v}_{e}(p_1) \gamma^{\mu} \omega_+ u_{e}(p_2)
          = \sqrt{s}\,(0, i,-1,0)~, \nn
\end{align} 
which is orthogonal with respect to the total initial-state momentum
$p_1\!+\!p_2=p_+\!+\!p_-$.

Exploiting the various properties of the polarization vectors and 
electron--positron currents, we end up with the following leading non-local
contributions to the right- and left-handed transverse amplitudes at high 
energies: 
\begin{eqnarray}
\label{MRL}
{\cal M}_R &\approx& (ie)^2\,\beta\sj\,\Bigl[\,\frac{s}{2}\,
                     \sigtwo^\prime(\MW^2)\Bigr]\,\frac{s}{D(s)}\,\bigg\{ 
                     \frac{s}{\cw^2}\,\sigthree(s)
                     + \MZ^2\,\Bigl[1+\sigfive(s)+\sigsix(s)\Bigr] \bigg\} 
                     \nn\\
{\cal M}_L &\approx& (ie)^2\,\beta\sj\,\Bigl[\,\frac{s}{2}\,
                     \sigtwo^\prime(\MW^2)\Bigr]\,\frac{s}{D(s)}\,\bigg\{ 
                     -\,\frac{s}{2\sw^2}\,\Bigl[1+\sigone(s)
                     -\frac{\sw^2}{\cw^2}\,\sigthree(s)\Bigr] \nn\\[1mm]
           &&        {}+ \MZ^2\,\Bigl[1+\sigfive(s)+\sigsix(s)\Bigr] \bigg\}~. 
\end{eqnarray}
A few remarks are in order here. First of all, the matrix elements 
${\cal M}_{R,L}$ given in \eqn{MRL} are valid only if both $W$ bosons have the 
same transverse helicity. The matrix elements vanish if the $W$ bosons have 
opposite helicity. Second, the factor $\,s\,\sigtwo^\prime(\MW^2)/2\,$ is the 
leading high-energy component of the non-local triple gauge-boson vertex of 
\eqn{TGC_nonlocal}, which has to be contrasted with the corresponding factor 
$-1$ for the local triple gauge-boson vertex. The derivative 
$\,\sigtwo^\prime(\MW^2)\,$ originates from the typical non-local expression
\begin{equation}
  \frac{\sigtwo(p_-^2)-\sigtwo(p_+^2)}{p_-^2-p_+^2} 
  \,\stackrel{p_{\pm}^2\to\MW^2}{-\!\!\!-\!\!\!-\!\!\!\longrightarrow}\, 
  \sigtwo^\prime(\MW^2)~.
\end{equation} 
Since $D(s)\propto s^2$ at high energies [see \eqn{D(s)}], ${\cal M}_L$ will 
have an incorrect high-energy behaviour, growing with $s$ as a result of the 
non-local triple gauge-boson factor $\,s\,\sigtwo^\prime(\MW^2)/2$. 
[Note that this is not the case for ${\cal M}_R$ in view 
of the fact that $\sigthree(s)$ vanishes at high energies.] 
On the basis of this observation we conclude that the \BBC approach has a
problem with high-energy unitarity.


\end{document}